\documentclass[letterpaper,twocolumn,10pt]{article}
\usepackage{CJKutf8}
\usepackage{usenix2019_v3}

\usepackage[english]{babel}
\usepackage{multirow}
\usepackage{booktabs}
\usepackage{supertabular}
\usepackage{pifont}
\usepackage{float}
\usepackage{graphicx}
\usepackage{underscore}
\usepackage{enumitem}
\usepackage{comment}
\usepackage{amsmath}
\usepackage{amssymb}
\usepackage{hyperref}
\usepackage{xurl}
\usepackage{subfig}
\usepackage{balance}

\begin{document}

\newcommand{\blue}[1]{\textcolor{blue}{#1}}
\newcommand{\red}[1]{\textcolor{red}{#1}}
\newcommand{\gray}[1]{\textcolor{gray}{#1}}

\newcommand{\ttt}[1]{\textbf{\texttt{\small{#1}}}}
\newcommand{\ttsmall}[1]{\textbf{\texttt{\footnotesize{#1}}}}
\newcommand{\ttquote}[1]{\texttt{`\texttt{#1}'}}

\date{}

\title{INFERMAL: Inferential analysis of maliciously registered domains} 

\author{
    Yevheniya Nosyk, Maciej Korczy\'nski*, Sourena Maroofi, Jan Bayer, \\ Zul Odgerel, Andrzej Duda\\
    {\normalsize KOR Labs / Université Grenoble Alpes \vspace{+0.2cm}} \\
    ICANN Org Project Coordination: \\
    Samaneh Tajalizadehkhoob*, Carlos Gañán \vspace{+0.2cm} \\
    {
    \normalsize * Scientific Contact: }\\ {\normalsize \texttt{maciej.korczynski@korlabs.io}} \\ {\normalsize \texttt{samaneh.tajali@icann.org}}
}

\maketitle
\begin{abstract}

Cybercriminals have long depended on domain names for phishing, spam, malware distribution, and botnet operation. To facilitate the malicious activities, they continually register new domain names for exploitation. Previous work revealed an abnormally high concentration of malicious registrations in a handful of domain name registrars and top-level domains (TLDs). Anecdotal evidence suggests that low registration prices attract cybercriminals, implying that higher costs may potentially discourage them. However, no existing study has systematically analyzed the factors driving abuse, leaving a critical gap in understanding how different variables influence malicious registrations. In this report, we carefully distill the inclinations and aversions of malicious actors during the registration of new phishing domain names. We compile a comprehensive list of 73 features encompassing three main latent factors:  registration attributes, proactive verification, and reactive security practices. Through a GLM regression analysis, we find that each dollar reduction in registration fees corresponds to a 49\% increase in malicious domains. The availability of free services, such as web hosting, drives an 88\% surge in phishing activities. Conversely, stringent restrictions cut down abuse by 63\%, while registrars providing API access for domain registration or account creation experience a staggering 401\% rise in malicious domains. This exploration may assist intermediaries involved in domain registration to develop tailored anti-abuse practices, yet aligning them with their economic incentives.

\end{abstract}

\section{Introduction}

Cybercriminals extensively exploit the Domain Name System (DNS) for a broad range of illegal or malicious activities including phishing, spam distribution, botnet command and control, or malware dissemination. Domain names serve as pathways that direct victims to servers hosting harmful content. Cybercriminals may either register domain names for malicious purposes or use legitimate domains registered by benign users that later fall prey to vulnerabilities in software, resulting in their exploitation for hosting malware or phishing websites. Furthermore, cybercriminals frequently abuse various services such as free subdomain providers~\cite{comar}, the InterPlanetary File System (IPFS)~\cite{ipfs}, and bucket storage services~\cite{bucket} to distribute malicious content using domain names associated with these legitimate services. 

The 2023 Phishing Landscape study~\cite{landscape} reveals that over 1.1 million unique domains were involved in criminal activities and added to blocklists between May 2022 and April 2023. It may seem that registering domain names for malicious purposes requires a significant financial investment by cybercriminals, potentially making it less attractive. However, it remains the most common approach---overall, cybercriminals deliberately registered 725~K blocklisted domains~\cite{landscape}. Once abusive domain names appear on blocklists, they are mitigated either by blocking communication at the network level (by ISPs, mail service operators, or DNS resolver operators) or, more effectively, at the DNS level by registrars or top-level domain (TLD) registries. Therefore, attackers have to fulfill a constant need for numerous single-use domain names to evade detection and maintain their malicious activities effectively~\cite{spamregistration}.

Malicious registrations are not uniformly distributed over different actors of the DNS ecosystem---they tend to be skewed toward certain domain name registrars~\cite{spamregistration, sunrise, dnsabuse2022study,whois-com,predator,registrations-in-eu} and TLDs~\cite{antivirus,dnsabuse2022study}. A notable example is Freenom, which previously managed five country-code TLDs (\texttt{.tk}, \texttt{.ml}, \texttt{.ga}, \texttt{.cf}, and \texttt{.gq}) and offered free registrations. As of 2013, Freenom accounted for 28\% of all malicious domain registrations~\cite{landscape}. Another observed trend among cybercriminals is the shift from legacy to new generic TLDs that tend to offer competitive prices (sometimes below \$1), which might attract attackers~\cite{sunrise}. 

A large body of existing work investigated the factors that make certain registrars or TLDs appealing to cybercriminals. Some studies observed that malicious domains are frequently registered during large campaigns~\cite{spamhaus-bulk,criminal-abuse,lifetimes} possibly using the registrars that facilitate bulk registrations. Others anecdotally suggested that low prices may play a role in malicious registrations~\cite{metrics,dnsabuse2022study,sunrise,registrar-level}. However, no study has thoroughly  analyzed the diverse range of factors that can change the attacker preference toward using certain registrars and TLDs. Moreover, existing data-driven studies have not identified specific measures that effectively increase barriers to DNS abuse while remaining appealing to legitimate clients.

In this report, we bridge this research gap with a thorough analysis of different factors that may influence the choice of certain registrars and TLDs by malicious actors when registering new phishing domains. We gather and analyze various TLD attributes / registrar practices to define 73 features encompassing three main groups of latent factors of DNS abuse: registration attributes (e.g., prices, payment methods, additional services), proactive verification (e.g., checking registrant data), and reactive security practices (e.g., uptimes). Finally, we develop two models: the first model aims to estimate the impact of the features on the number of maliciously registered domains, while the second one indicates whether the registrar or TLD level features are favored by attackers alone or also by legitimate users. 

Overall, we make the following contributions:

\begin{itemize}
    \item We introduce a statistical regression model to analyze the relationship between various features and the concentration of phishing domain names using \textit{registrar-TLD} pairs as the unit of analysis. 
    This is motivated by the fact that certain variables are inherited from TLD registry practices, while others directly originate from registrars.
    
    \item We find that domain abuse is closely linked to discounts, with each dollar off leading to a 49\% increase in malicious registrations. Free services like web hosting result in an 88\% rise in phishing domains, while stringent restrictions reduce abuse by 63\%. Registrars offering API access for domain registration or account creation see a 401\% increase in malicious domains. Mitigation times have little impact, likely because even brief uptimes may provide phishers with valuable credentials and financial gain.
    
    \item We create a  statistical model to identify features preferred by malicious versus benign domain registrations. Using fine-grained regression modeling at the \textit{domain level}, we analyze how various factors influence whether a domain is classified as malicious or benign.

    \item The analysis shows that discounts attract more malicious actors than legitimate registrants, while various restrictions reduce the likelihood of a domain being maliciously registered by about 19\%. This suggests that malicious users are more likely than legitimate ones to consider these restrictions when selecting a TLD and a registrar.
    
\end{itemize}

The rest of this report is organized as follows. Section~\ref{sec-background} provides the background on domain name registrations and anti-abuse measures. We review the related work in Section~\ref{sec-related-work}. Section~\ref{sec-datasets} presents the datasets we use to collect the  set of features discussed in Section~\ref{sec-features}. We analyze the malicious registrations in Section~\ref{sec-results} and evaluate the driving factors of abuse in Section~\ref{sec-driving-factors}. Section~\ref{sec:discussion} discusses the lessons learned while Section~\ref{sec:conclusions} concludes the paper. 

\section{Background\label{sec-background}}

This section provides background on the Registrant -- Registrar -- Registry (RRR) model, DNS abuse, anti-abuse measures, and malicious phishing domains.

\subsection{DNS Ecosystem: 
Registries, Registrars, and Registrants}

The Domain Name System (DNS) is one of the fundamental components of the modern Internet providing the mapping between human-readable domain names (e.g., \texttt{example.com}) and IP addresses (e.g., \texttt{198.51.100.0}). Each top-level domain (TLD), e.g., \texttt{.com} (legacy generic TLD), \texttt{.top} (new generic TLD), or  \texttt{.de} (country-code TLD),  is managed by a \textit{registry}---an organization that sets the registration terms and prices, maintains the DNS zone file, and configures DNSSEC. As of July 2024, the DNS Root Zone Database contains 1,591 top-level domains~\cite{root-db}.

Registries typically delegate the responsibility of selling domain names to \textit{registrars} that set up contractual agreements with registries and sell domain names under the relevant TLDs. Obtaining ICANN accreditation is essential for selling gTLD domain names, whereas for ccTLDs, accreditation from local registry operators may suffice (e.g., SIDN for \texttt{.nl} domains \cite{sidn}). Becoming an ICANN-accredited registrar involves a rigorous process including meeting eligibility criteria, signing agreements, and paying fees. Successful applicants must demonstrate financial stability, technical capability, and business expertise, while committing to ongoing compliance with ICANN policies and procedures~\cite{statement}. 

Finally, a \textit{registrant} is any entity (benign or malicious) that registers a domain name and agrees to the registrar terms providing accurate personal information as required by the registry, ICANN, or both.

\subsection{DNS Abuse}

Cybercriminals extensively leverage the DNS and domain names for a wide panoply of illegal and malicious activities. In 2019, a group of domain registries and registrars, including GoDaddy~\cite{godaddy-home}, Tucows~\cite{tucows-home}, Namecheap~\cite{namecheap}, Public Interest Registry (\texttt{.org})~\cite{pir-home}, Neustar (\texttt{.biz}, \texttt{.us})~\cite{neustar-registry} and Afilias (now part of Identity Digital)~\cite{afilias-registry} voluntarily created the DNS Abuse framework to fight against DNS abuse~\cite{dnsabuseframework}. The framework aimed to provide a clear definition of DNS abuse and establish guidelines for registries and registrars to combat DNS abuse more effectively and consistently across the industry. In particular, they categorized DNS abuse into 5 different types: malware, botnets, phishing, pharming, and spam (when used to distribute the other threats). These activities exploit the DNS infrastructure as a delivery mechanism for their illicit operations~\cite{inj, icann-abuse}.

Abuse handling policies and procedures vary among registrars and the operators of generic TLDs and ccTLDs. ICANN-accredited registrars must adhere to specific abuse handling guidelines. Previously, they were expected to address abuse complaints but specific requirements and timelines for response were not clearly defined. The focus was primarily on reactive abuse measures. As of April 2024, gTLD registries are also required to proactively address abuse due to new contractual amendments~\cite{amendment}. In contrast, ccTLDs are considered national resources with unique characteristics and do not have contractual agreements with ICANN for abuse-handling policies. Thus, their procedures depend on local regulations and the voluntary practices of individual registries and registrars.

\subsection{Anti-Abuse Measures} 

TLD registries and registrars undertake various measures to prevent and mitigate abusive domain names, categorized into proactive verification, reactive security practices, and registration attributes. While not strictly preventive, registration attributes may indirectly deter abuse. 

The primary objective of proactive measures is to prevent malicious registrations from occurring. Certain TLDs are restricted to specific regions (e.g., \texttt{.eu}) or professions (e.g., \texttt{.abogado}). Some TLD registries and registrars implement identity verification processes known as Know Your Customer (KYC)~\cite{eurid-kyc} (e.g., \texttt{.dk}~\cite{dk-terms}). Additionally, other registries use machine learning techniques to identify suspicious domain names during registration (e.g., \texttt{.eu}~\cite{premadoma}, \texttt{.nl}~\cite{regcheck}, \texttt{.be}~\cite{dnsbelgium}), ensuring they are flagged before being added to the namespace and delegated to zone files. Another proactive strategy is to block the registration of domains containing keywords associated with well-known brands during the general availability period of TLDs (e.g., two registrars prevent the registration of such domains as discussed in Section~\ref{sec:trademark}).

When a domain name is involved in phishing or malware distribution, the TLD registry or registrar may take action to remove it at the DNS level if there is sufficient evidence of abuse and no legitimate content is being served. However, if the domain name itself is legitimate but vulnerable software on the site has been exploited~\cite{comar}, the issue cannot be resolved at the DNS level. Instead, the abuse must be handled by the hosting provider or webmaster, depending on whether the hosting is managed or unmanaged~\cite{Tajalizadehkhoob17a}. Taking action at the DNS level in such cases could cause collateral damage to the website visitors and the owners of benign domain names.

At the DNS level, several actions can be taken against abusive domains: i) the domain can be deleted from both the DNS zone and the namespace, ii) the domain can be delisted (suspended) from the DNS zone but remains in the namespace, iii) the domain can stay in both the zone and namespace, but its authoritative nameserver can be changed to a dedicated one, managed by the TLD registry specifically for this purpose.

Note that the diverse domain registration attributes proposed by registrars such as varying prices, bundled services (e.g., web hosting), and multiple payment methods (credit cards, PayPal, cryptocurrencies) could act as preventive measures. 
These features may either deter or, conversely, attract attackers seeking to register abusive domain names with particular registrars and TLDs.  
For example, GoDaddy recently updated its Terms of Use, requiring customers to have 50 or more domains in their accounts to use the Availability API~\cite{godaddy-api,godaddy-api-old}. This change could potentially impact the use of the GoDaddy API for malicious purposes, though its effect on abuse rates is yet to be determined.  

\subsection{Maliciously Registered Phishing Domains}

This study examines phishing abuse and domains registered with malicious intent. Phishing is widely recognized as a significant cyber threat and a prevalent form of DNS abuse. According to the 2023 annual report from the FBI, phishing is the leading type of digital crime, with over 300,000 complaints and losses exceeding \$160 million \cite{FBIreport23}.

We specifically focus on phishing because malware delivery URLs are less common and readily detected, spam domains do not always qualify as DNS abuse, and phishing typically provides clear evidence, such as screenshots of fraudulent sites.

\begin{figure}[t]
    \centering
    \includegraphics[width=0.5\textwidth]{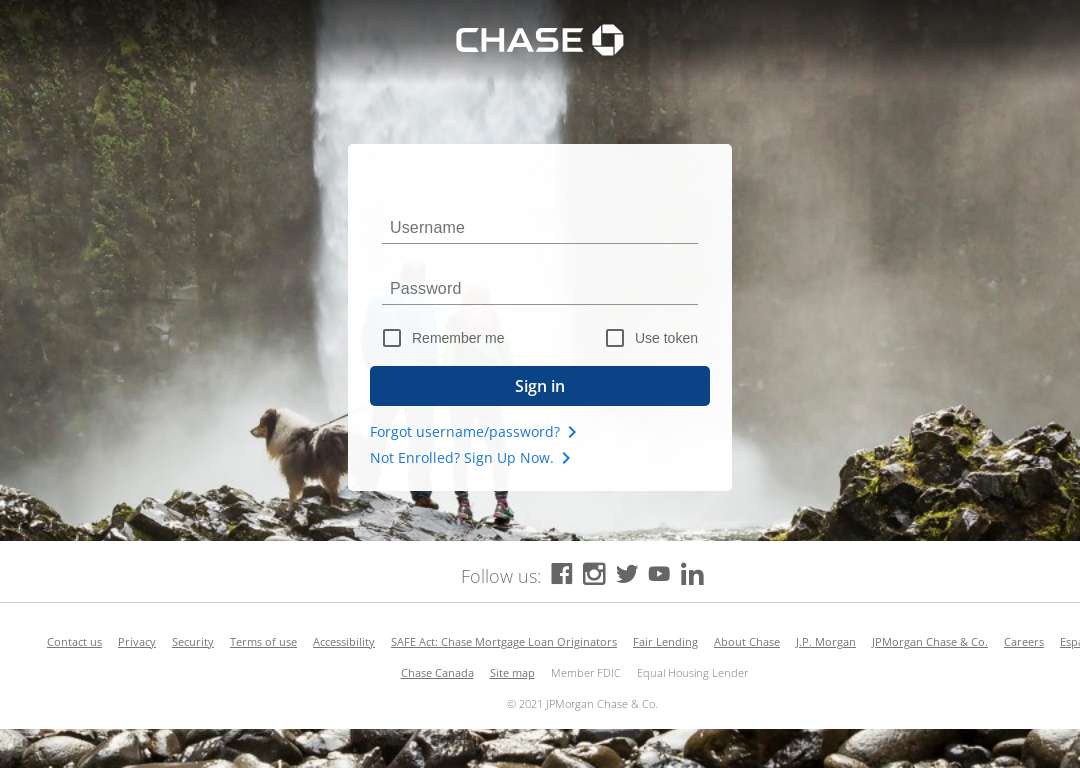}
    \caption{A phishing website impersonating Chase Bank (\url{hxxps://chase03.com/login}) detected by PhishTank on June 5, 2024. The domain contains the deceptive keyword ``chase,'' and was registered just one day prior, suggesting malicious intent.}
    \label{fig:chase-mal}
\end{figure}

\begin{figure}[t]
    \centering
    \includegraphics[width=0.5\textwidth]{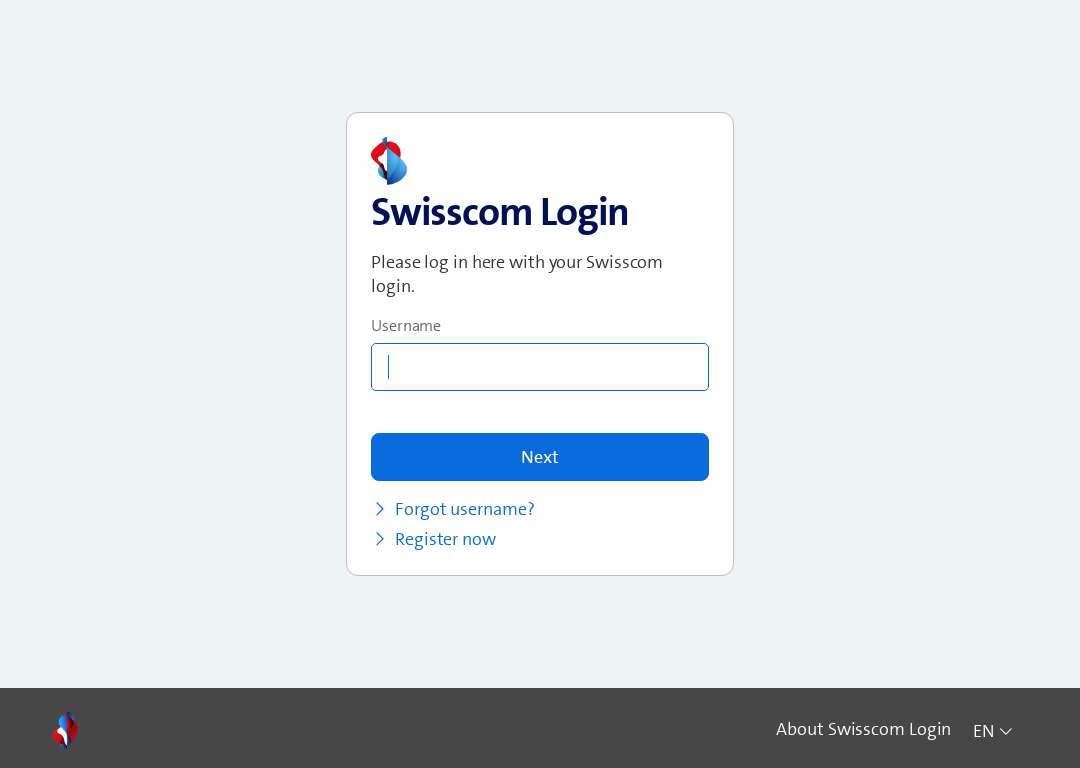}
    \caption{A phishing website impersonating Swisscom telecommunications provider (\url{hxxps://landpmullen.co.uk/wp-includes/widgets/rechnung/?token=}) detected by PhishTank on July 14, 2024. The domain, registered on October 28, 2008, contains a malicious URL with ``wp-content'' while the registered domain name features benign content. This suggests that the domain is benign but later compromised through vulnerabilities in the WordPress CMS.}
    \label{fig:swiss-comp}
\end{figure}

\begin{figure}[t]
    \centering
    \includegraphics[width=0.5\textwidth]{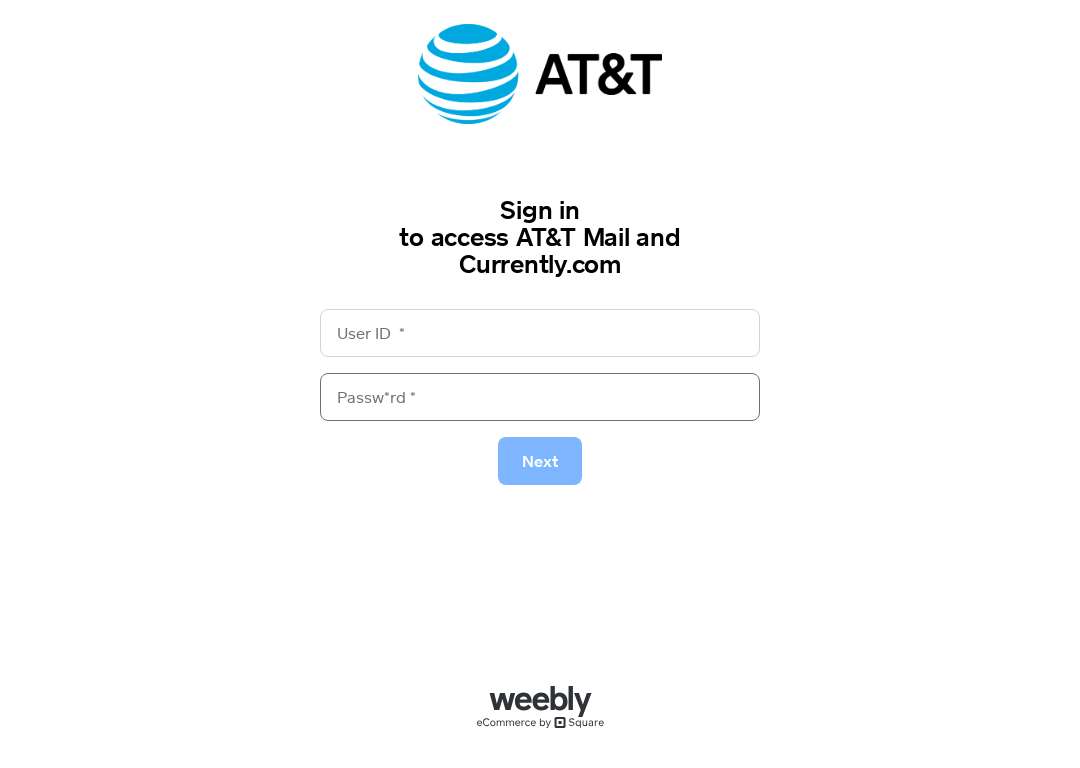}
    \caption{A phishing website impersonating AT\&T  telecommunications holding company (\url{hxxps://attsecure0.weebly.com}) detected by the APWG on July 22, 2024. The site exploits \url{weebly.com}---a free website builder and subdomain provider.}
    \label{fig:att-free}
\end{figure}

While some phishing domains are registered with purely malicious intent (or ``attacker-owned''~\cite{DeSilva2021}), others are  benign but may become compromised through, e.g. vulnerabilities in their content management systems (CMS)~\cite{comar}, etc. Attackers may also exploit free services, such as subdomain providers, to disseminate malicious content. Current phishing detection methods identify indicators of ongoing attacks, often conflating maliciously registered and compromised domains into common URL blocklists. Therefore, previous research has proposed methods to distinguish between these two groups~\cite{LePage2019,comar,DeSilva2021}. In the Appendix, we provide examples of three distinct types of phishing attacks: those involving maliciously registered domain names (Figure \ref{fig:chase-mal}), compromised websites (Figure~\ref{fig:swiss-comp}), and legitimate subdomain provider services (Figure~\ref{fig:att-free}). In this report, we focus on domains maliciously registered for phishing purposes, rather than benign ones that are later exploited.

\section{Related Work\label{sec-related-work}}

A large body of research focused on dissecting the registration patterns of attackers, understanding their preferences, and evaluating the response of various intermediaries to reported DNS abuse.

Previous research collected substantial evidence that malicious actors heavily use bulk domain name registration. Felegyhazi et al.~\cite{potential-blacklisting} discovered registration clusters from a small seed of known malicious domains. Subsequently, 73\% of newly inferred domains would eventually appear on blocklists. Later, in 2013,  Hao et al.  \cite{spamregistration} showed that 80\% of spam domains were registered in groups, 10\% belonging to batches of more than 200 registrations. Similar findings were observed at the \texttt{.eu} ccTLD for which 80\% of malicious registrations were associated with 20 campaigns~\cite{registrations-in-eu}. Moreover, Affinito~et~al.~\cite{lifetimes} examined two malicious domain registration spikes and noted that the two registrars behind offered bulk registration to their customers. Yet, Spooren~et~al.~\cite{premadoma} argued that attackers can evade the prediction models relying on bulk registration patterns if domain registrations are done on multiple days with periods of inactivity. 

It was also speculated that domain pricing plays a key role when choosing a particular registrar or a registry. Korczy\'{n}ski~et~al.~\cite{metrics} benchmarked top-level domains and found that TLDs offering domains for free contain 331 times more phishing domains than those charging a registration fee. Bayer~et~al.~\cite{dnsabuse2022study} observed similar trends when analyzing phishing counts across ccTLDs. They found that Freenom, the operator known to give out domains for free operated 5 out of 10 most abused ccTLDs. In a different study, Korczy\'nski~et~al.~\cite{sunrise} noticed a shift in abuse from legacy towards new gTLDs. The anecdotal evidence suggested that low pricing under certain new gTLDs might attract malicious actors, but the authors did not systematically prove this hypothesis due to the lack of  pricing data. More broadly, Liu~et~al.~\cite{registrar-level} showed that when CNNIC (\texttt{.cn} registry) tightened requirements to be fulfilled by registrants and increased the minimum domain price from \$0.12-0.15 to \$10, spammers switched to other TLDs.

Previous studies have examined patterns in domain registration associated with malicious activities and the steps taken by registrars to mitigate these threats. Vissers~et~al.~\cite{registrations-in-eu} analyzed 14 months of registrations at \texttt{.eu} ccTLD and noticed that during three malicious registration campaigns, attackers provided non-existing combinations of street addresses and countries. Phishers may use deceptive keywords, like famous brands. For instance, a recent study reported that Namecheap had blocked domains containing the keyword ``facebook''~\cite{dnsabuse2022study}.

Another factor that enters into play is how registrars and registries deal with abuse complaints. Liu~et~al.~\cite{registrar-level} showcased an example of a successful collaboration between eNom and LegitScript. The registrar took down all the domain names hosting rogue pharmacy domains. In a more recent study, Cheng~et~al.~\cite{internet-entities} analyzed the response of various Internet entities within China to reported gambling and adult content websites. Most registrars suspended abusive domains within 24 hours by either setting domain status codes to \textit{Client-Hold} or using sinkhole nameservers. More generally, Korczy\'{n}ski~et~al.~\cite{metrics}  found no correlation between the abuse response times and the prevalence of abuse, suggesting that abuse response times may not significantly affect the prevalence of abuse. 

Our report advances existing research in several significant ways. First, we clearly distinguish between compromised (websites) and maliciously registered domains, focusing exclusively on the latter for the analysis. Second, we compile a comprehensive set of features encompassing registration attributes,  proactive and reactive security measures. Finally, we propose two statistical models to empirically study the factors that influence the registration of malicious domains from the perspective of attackers.

\section{Datasets\label{sec-datasets}}

This section discusses the principal datasets used in the paper. We first present the TLD-List, our primary source of registrar and TLD features. We then curate a list of malicious domain names that appeared on blocklists deliberately registered by phishers. Finally, we create a list of benign domain names and carefully sample benign registrations to make the two datasets comparable. 

\subsection{TLD-List}

The TLD-List service~\cite{tldlist} has been collecting data on TLDs and registrars since 2015. With a unique focus on pricing, it includes domain registration costs, discounts, and free features. Pricing is being updated every three hours, and other details are verified weekly. We subscribed to this service, collecting daily snapshots with data on the registrar payment methods, free features (e.g., SSL/TLS certificates), and prices. 

Overall, the collected datasets span 75 domain name registrars and more than 1,500 top-level domains. They enable us to verify the prices and services offered by registrars on the registration day of a specific  domain name.

We validated the data by randomly sampling 20 registrar/TLD pairs across different dates during the analysis period and manually verifying the dataset. While all data was generally accurate, we found 7 discrepancies related to payment methods. However, since we aggregate payment methods in our model (see Section \ref{sec:feature-eng}), these minor inconsistencies do not affect our findings.

While our primary source, the third-party registration data, has been validated and proved to be highly accurate, it may not fully reflect the attributes actually chosen by registrants of both malicious and benign domains. Some features, like certain restrictions, are defaults, while others may be used optionally rather than  specifically selected. Without comprehensive ground truth data, we cannot estimate the extent of any potential bias in this assumption. 

\subsection{Maliciously Registered Phishing Domains}

To understand the registration preferences of phishers, we analyze the domain names that satisfy two conditions: i) they were involved in phishing activities and ii) they were deliberately registered by cybercriminals. 

We first collect 534~K blocklisted URLs from three phishing feeds maintained by the Anti-Phishing Working Group (APWG)~\cite{apwg}, PhishTank~\cite{phishtank}, and OpenPhish~\cite{openphish}, spanning the period between August 2023 and January 2024. We selected the feeds because they have been commonly used in prior research~\cite{comar,sunrise, metrics,daar}, and are maintained by reputed organizations. We process all the URLs and extract 108~K registered domains, noting that some are benign but have been abused by malicious actors. We begin by excluding the domains associated with URL shorteners (e.g., \texttt{bit.ly})~\cite{urlshortener} and subdomain providers (e.g., \texttt{000webhostapp.com})~\cite{subdomain-providers}, known to be used for delivering malicious content~\cite{le2018using,nikiforakis2014stranger,le2015security} but registered for legitimate purposes. 

Next, we perform a set of measurements for all the blocklisted domains during one month after being reported. Specifically, we retrieve registration data (using WHOIS or RDAP protocols) and DNS \texttt{A} records. While compromised domains should only have the malicious content removed, maliciously registered ones should result in a takedown action at the DNS level as evidenced by the \texttt{NXDOMAIN} DNS response code and the Extensible Provisioning Protocol (EPP) status code set to \texttt{clientHold} or \texttt{serverHold} ~\cite{RFC5731}. Our analysis includes only domains that were mitigated at the DNS level within the one-month monitoring period. 

Finally, given that maliciously registered domain names are often used for malicious activities shortly after registration~\cite{spamregistration}, we only include the domain names registered within 90 days prior to being blocklisted. While this strict approach may still miss some malicious registrations, it helps ensure that we do not include compromised domains.

We apply these heuristics to a set of 108~K domains, resulting in the classification of 28~K domains registered maliciously, spanning 157 registrars. Table~\ref{registrars-domains-all} in the Appendix shows the 20 most abused registrars, accounting for 83.8\% of maliciously registered domains in our dataset. 

To associate maliciously registered domain names with the daily-collected registration features, we initiate a WHOIS scan immediately after blocklisting. If the domain remains active, we extract the registrar IANA ID and the registration date.  We exclusively consider domain names for which we have registrar features, referencing the list of registrars supported by the TLD-List dataset. As our analysis operates at both the registrar-TLD and domain name levels rather than at the registrar level, we argue that our sample of maliciously registered domain names serves as a representative basis for analyzing attacker preferences. 

Overall, we obtained the list of 14,474 maliciously registered domains spread across 165 TLDs and 31 registrars.  Table~\ref{registrars-tlds-malicious} shows 20 most frequently observed registrar/TLD pairs in our dataset. Furthermore, Table~\ref{tlds} in the Appendix presents the top 20 TLDs while Table~\ref{registrars-malicious} displays the number of domain names per registrar. 

\begin{table}[t]
    \scriptsize
    \centering 
    \caption{20 most frequently observed registrar/TLD pairs in our dataset of maliciously registered domain names.     \label{registrars-tlds-malicious} } 
    \begin{tabular}{clcc}
        \toprule
        \textbf{Rank} & \textbf{Registrar} & \textbf{TLD} & \textbf{\#Domains}\\
            \midrule
            1. & NameSilo & \textbf{\texttt{\scriptsize{top}}} & 1,807 \\
            2. & NameSilo & \textbf{\texttt{\scriptsize{com}}} & 852 \\
            3. & GoDaddy & \textbf{\texttt{\scriptsize{com}}} & 832 \\
            4. & Hostinger & \textbf{\texttt{\scriptsize{online}}} & 764 \\
            5. & NameSilo & \textbf{\texttt{\scriptsize{info}}} & 513 \\
            6. & Hostinger & \textbf{\texttt{\scriptsize{com}}} & 479 \\
            7. & Namecheap & \textbf{\texttt{\scriptsize{com}}} & 479 \\
            8. & Alibaba Cloud & \textbf{\texttt{\scriptsize{com}}} & 327 \\
            9. & NameSilo & \textbf{\texttt{\scriptsize{xyz}}} & 233 \\
            10. & Hostinger & \textbf{\texttt{\scriptsize{cloud}}} & 225 \\
            11. & NameSilo & \textbf{\texttt{\scriptsize{buzz}}} & 222 \\
            12. & Sav & \textbf{\texttt{\scriptsize{com}}} & 211 \\
            13. & Alibaba Cloud & \textbf{\texttt{\scriptsize{shop}}} & 197 \\
            14. & NameSilo & \textbf{\texttt{\scriptsize{us}}} & 191 \\
            15. & Hostinger & \textbf{\texttt{\scriptsize{site}}} & 179 \\
            16. & NameSilo & \textbf{\texttt{\scriptsize{life}}} & 178 \\
            17. & NameSilo & \textbf{\texttt{\scriptsize{sbs}}} & 171 \\
            18. & Hostinger & \textbf{\texttt{\scriptsize{shop}}} & 156 \\
            19. & NameSilo & \textbf{\texttt{\scriptsize{cc}}} & 149 \\
            20. & Alibaba Cloud & \textbf{\texttt{\scriptsize{top}}} & 148 \\
        \bottomrule
    \end{tabular}
\end{table}

\subsection{Benign Domains}

Some factors that attract attackers, like competitive pricing or free features, may also appeal to legitimate users. To understand the differences, we curated a list of benign domain names as a baseline.

We first gather all the registered domain names appearing in the Centralized Zone Data Service~\cite{icann-czds}, Google Certificate Transparency logs~\cite{calidog}, and zone files of \texttt{.se}, \texttt{.nu}, \texttt{.ch}, \texttt{.li} (all retrieved via AXFR zone transfers). We then perform a WHOIS scan of them to get registration dates as well as IANA IDs, keeping only the domains created during the same time window as the maliciously registered ones. We further remove 1~M domains appearing in Spamhaus~\cite{dbl} and SURBL blocklists~\cite{surbllists} and keep the list of 19~M domains created at registrars supported by the TLD-List dataset. 

To make the two datasets comparable, we need a representative sample of benign domains that takes into account registrar market shares. We refer to the ICANN Monthly Registry Reports~\cite{icann-report} in which each gTLD registry gives the number of domain names managed by each registrar under a particular gTLD. Although these numbers exclude ccTLD domains, they can still serve as an estimate of the registrar market share. We then use the obtained ratios to perform the stratified sampling of 19~M benign domains. We finally collect all the registration and proactive features, which result in a dataset of 15.4~K domains under 259 TLDs originating from 38 registrars. 

\section{Features\label{sec-features}}

In this section, we delve into the preliminary analysis of the registration attributes and anti-abuse practices put forth by domain name registrars and registries. We categorize them into three distinct categories: i) registration attributes, ii) proactive verification, and iii) reactive security practices. Throughout the remainder of this section, we provide their overview and the rationale for choosing them.  

\subsection{Registration Attributes}

We first describe the pre-selected registration attributes, most of which are derived from the TLD-List dataset. For any missing information not available in these datasets, we manually collect the necessary data. Table~\ref{tab:features-registration} in Appendix provides the summary of the collected features:

\textbf{Free API:} the registrar APIs enable users to search, purchase, and manage domains, allowing cybercriminals  to fully automate the setup of malicious infrastructures. The boolean \textit{free_api} feature indicates whether registrants can access the API without any prerequisites such as a reseller account or a paid subscription. We also define the boolean features \textit{api_create_account} for the account creation and \textit{api_register_domain} for domain registration.

\textbf{Free bulk search:} registering multiple domain names at once may help attackers maintain resilience against quick blocklisting and enables running concurrent campaigns. Research indicates that malicious domains are often registered in batches~\cite{spamhaus-bulk, criminal-abuse, lifetimes}. Therefore, we examine the capability to search domains in bulk (numerical \textit{free_bulk_search_number}) and any associated discounts (boolean \textit{bulk_discount}).

\textbf{Available payment methods}: malicious actors prioritize anonymity often opting for payment methods  harder to trace such as cryptocurrency. For instance, ransomware operators predominantly use Bitcoin to receive payments from victims \cite{ransomware}. We define 24 boolean features for each payment method in the TLD-List dataset including PayPal, Bitcoin, and others (see Features \S12 - \S35 in Table~\ref{tab:features-registration} in the Appendix).

\textbf{Estimated prices:} existing research suggests that pricing significantly influences the registration preferences of attackers. They tend to favor domain name registrars and TLDs that offer the most competitive rates. We establish three numerical features: \textit{price\_register}, \textit{price\_renewal}, and \textit{price\_transfer}, denoted in \$.

\textbf{Discounts:} discounts on domain registrations may attract attackers. For instance, one registrar offers lower prices for bulk registrations of 50 or more domains. The discounts, which vary by TLD, can reduce the cost of building malicious infrastructure. We define three numerical features in \$ to capture the discounts: \textit{discount_register}, \textit{discount_renewal}, and \textit{discount_transfer}.

\textbf{Pricing terms:} certain registrars impose specific conditions on domain purchases. For example, discounted pricing might apply only to a limited number of domains or require purchase through an affiliate link. We define these conditions and purchase types using boolean features (\textit{term_new_customer_only_register} and \textit{term_new_customer_only_transfer}) and numerical features (\textit{term_limit_per_customer_register} and \textit{term_limit_per_customer_transfer}).

\textbf{Free web hosting:} previous research indicated that attackers typically do not invest significant effort in creating fully functional websites~\cite{comar}. They may leverage free hosting plans to host basic content on newly registered (malicious) domain names. If such a service is included for free in each domain registration, we set \textit{free_web_hosting} to True. 

\textbf{Free SSL/TLS certificates:} as of August 2020, 77.6\% of phishing websites used SSL/TLS certificates~\cite{trends}. Attackers may value free certificates for malicious domains despite the risk that they appear in Google Certificate Transparency logs, thereby increasing the chances of phishing detection~\cite{CTLogs}. We examine the impact of free certificates on the registrar selection using the boolean \textit{free_ssl_cert} feature.

\textbf{Free email}: registrars 
may offer free email boxes and/or email forwarding to their registrants. This service may also be exploited by attackers to deliver malicious content to their victims such as phishing links. Thus, we introduce two boolean features: \textit{free_email_account} and \textit{free_email_forward}.

\textbf{Free DNS service}: registrars commonly offer customers a free DNS service, effectively eliminating the need to establish and maintain a custom authoritative nameserver infrastructure. Spam domain owners~\cite{spamregistration} benefit from such a service as it reduces the overhead required to set up operational domain names. We define the \textit{free\_dns} boolean feature.

\textbf{Free DNSSEC signing:} registrars offering DNS services may provide free cryptographic signing of domain names, which may boost the domain reputation, even if not directly relevant to phishing. The \textit{free_dnssec} feature is set to True if the registrar signs the domain without requiring clients to upload custom \texttt{DS} records.

\textbf{WHOIS privacy price}: WHOIS/RDAP services reveal domain registration data, which can expose malicious actors~\cite{registrations-in-eu}. The General Data Protection Regulation (GDPR)\cite{gdpr} mandates masking personal details of European Economic Area (EEA) registrants, and some registrars apply this to non-EEA registrations, sometimes for free\cite{whowas}. Despite GDPR, we include the \textit{price_whois_privacy} feature in our analysis.

\subsection{Proactive Verification}

To assess proactive measures, we actively create registrant accounts and add various borderline domain names to a cart, empirically testing the presence of proactive security practices prior to the domain purchase. We review below the examined features  (Table~\ref{tab:features-proactive-reactive} in Appendix provides their summary):

\textbf{Syntactic validation of the registrant personal information:} The ICANN SSAC Report on Domain Name Registration Data Validation~\cite{icann-validation} outlines three validation types: syntactic, operational, and identity. We test 38 registrars by attempting to create accounts with syntactically incorrect data, including: i) an email address missing the ``@'' symbol, ii) a phone number exceeding 15 digits~\cite{tel-number}, and iii) a physical address with a 3-digit postal code, invalid for the selected country. We then define three boolean features to indicate whether registrars accept this incorrect data without warnings: \textit{email_syntactically_validated}, \textit{phone_syntactically_validated}, and \textit{address_syntactically_validated}.

\textbf{Operational validation of the registrant information:} the Registrar Accreditation Agreement~\cite{raa} mandates that ICANN-accredited registrars collect accurate contact information, while a recent European Commission directive requires verification of at least one contact method~\cite{ec-cysec}. Although ICANN allows verification within 15 days post-registration~\cite{accuracy-obligation}, our focus is on proactive verification. We test whether registrars verify contact email addresses and phone numbers during account creation or before domain purchase. By providing our genuine contact details, we expect verification through email or SMS. The features \textit{email_operational_validated} and \textit{phone_operational_validated} indicate if such verification is performed.

\textbf{Domain registration warnings and restrictions:} certain domain names may trigger suspicion during registration if they include well-known brand names or random character sequences. Registrars may issue warnings or block these domains. We define three labels for such scenarios: i) \texttt{a9e86e6d5d4c676441da} (the first 20 characters of the MD5 hash of ``DNS abuse''), ii) \texttt{office365-my-account}, and iii) \texttt{facebook-login-page}. The latter two are among the most targeted brands in our dataset of 534~K phishing URLs. For each registrar-TLD pair, we attempt to add these domains to the cart and proceed through all steps until prompted for payment. If succeeded, we set the corresponding boolean features to True: \textit{random\_warning}, \textit{random\_prevention}, \textit{office365\_warning}, \textit{office365\_prevention}, \textit{facebook\_warning}, \textit{facebook\_prevention}. We do not complete the purchase to avoid any potential brand infringement issues.

\textbf{Registration restrictions:} certain registries verify registrants' identities to ensure compliance with local regulations and to enhance the overall security of their domain ecosystem (e.g., \texttt{.dk}~\cite{dk-terms}). For instance, when the CNNIC mandated formal documentation and validation for individual registrations, it significantly reduced spam domains under the \texttt{.cn} TLD \cite{registrar-level}. Intuitively, attackers would avoid such TLDs and registrars. However, if these practices were implemented globally, malicious actors might adapt by resorting to identity theft for fraudulent registrations or compromising legitimate websites. We define 14 boolean features related to restrictions: (see Features \S12 - \S25 in Table~\ref{tab:features-proactive-reactive} in the Appendix). 

\subsection{Reactive Security Practices}

To evaluate reactive security practices at the TLD-registrar level, we measure mitigation times and, for a subset of our data, use the Netbeacon Reporter\footnote{\url{https://netbeacon.org/reporting}} to notify the most appropriate registrars for effectively mitigating abuse at the DNS level.

\textbf{Malicious domain name uptimes:} successful domain name registration is not the ultimate goal for attackers---they must remain operational to profit. For each unique abusive domain name, we measure its uptime (or persistence of abuse~\cite{metrics}). Uptime is defined as the duration between the blocklisting of a malicious URL and the mitigation of abuse at the DNS level. We measure mitigation by checking if the \texttt{A} record query returns \texttt{NXDOMAIN} or if WHOIS shows that the domain has been placed on hold by the registry (EPP \texttt{serverHold} status) or the registrar (EPP \texttt{clientHold} status). 

Initially, we measure uptime at the instant of acquiring the malicious URL followed by repeated measurements over the next month (approximate times): at 5 minutes, 15~m, 30~m, 1 hour, 2~h, 3~h, 4~h, 5~h, 6~h, 12~h, 24~h, 36~h, and 48~h after blocklisting, and then every 12 hours thereafter. Since phishing attacks are typically mitigated within the first day after blocklisting \cite{dnsabuse2022study}, we perform more frequent scans initially and less frequent scans later on. Some URLs from blocklists are already mitigated at the time of the first scan. In these cases, we calculate the time between blocklisting and the first measurement. This period is usually very short and provides a good approximation of the mitigation time. We calculate a median uptime at the TLD-registrar level and create a numerical \textit{uptime\_not\_notified} feature.

\textbf{Malicious domain name uptimes with notifications:} for a subset of maliciously registered domain names, notifications are sent to registrars via Netbeacon Reporter during the initial measurement, using abuse contact information extracted from WHOIS/RDAP records. We then calculate the median uptime (represented by the \textit{uptime_notified} feature) at the TLD-registrar level. 

\section{Descriptive Analysis of Features\label{sec-results}}

After collecting the features, we analyze the registration, proactive, and reactive security measures used by registrars and TLD registries, focusing on 14.5~K malicious domains deliberately registered by attackers.

\begin{figure}[t]
    \centering
    \begin{minipage}[b]{0.48\columnwidth}
        \includegraphics[width=\textwidth]{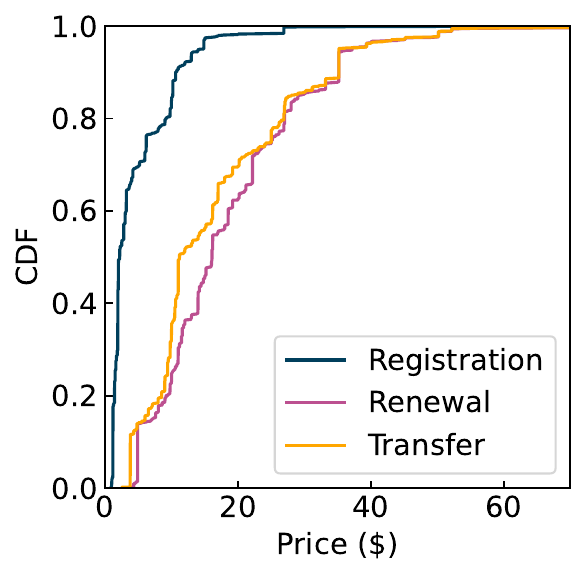}
        \caption{Distribution of registration, renewal, and transfer prices (in \$) proposed by registrars regarding malicious domains.}
        \label{fig:prices}
    \end{minipage}
    \hfill   
    \begin{minipage}[b]{0.48\columnwidth}
        \includegraphics[width=\textwidth]{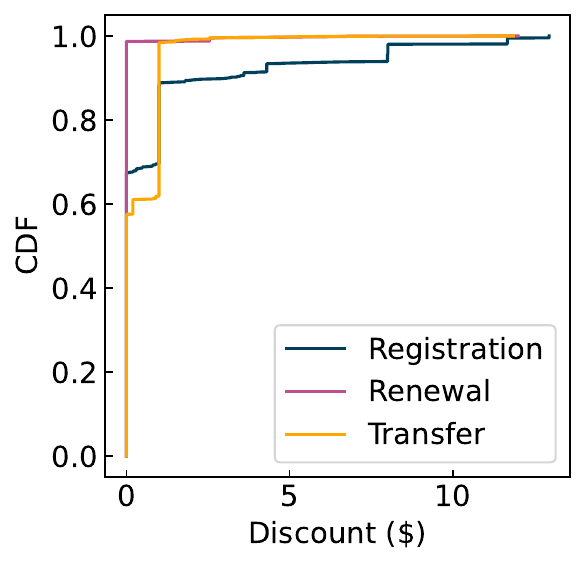}
        \caption{Distribution of registration, renewal, and transfer discounts (in \$) proposed by registrars regarding malicious domains.}
        \label{fig:discounts}
    \end{minipage}
\end{figure}

\subsection{Prices, Discounts, and Fees}

We start by analyzing the cost of domain name registration, renewal, and transfer. Our core assumption is that malicious actors are drawn to lower prices, particularly when discounts or special offers are available, as these reduce the overall cost of establishing malicious infrastructures.

Figure~\ref{fig:prices} shows the distribution of registration, renewal, and transfer prices for 14.5~K domains. Registering a domain is typically cheaper than transferring or renewing it. Since malicious domains generally have short lifespans, attackers are less concerned with transfer or renewal costs. Registration prices range from \$0.78 to \$69, with nearly 50\% costing \$2 or less. Examples of expensive domains include \texttt{usps.bar} at \$69, \texttt{support-fb.sh} at \$59.99, and \texttt{dhlcenter.net} at \$56. We hypothesize that  while attackers generally prefer cheaper options, cost may become less of a concern when they have access to a large supply of stolen credit cards or cryptocurrencies.

Registrars may also offer various forms of discounts, either by deducting a fixed amount or a percentage from the regular price. Figure~\ref{fig:discounts} shows the distribution of discount amounts for registration, renewal, and transfer. Most advertised prices lack promotions, especially for renewals. Typically, new registrants are attracted with lower registration prices but pay full rates upon renewal. Discounts for registration and transfer range from \$0.01 to \$12.95.

The presented registration, transfer, and renewal prices may also be subject to various terms, as was the case for 7,168 (49.52\%) domain names. We hypothesize that attackers are particularly sensitive to registration prices, as lower costs may enable them to purchase large numbers of domains at once. In particular, 4,423 reduced registration prices were valid for one domain only. Two registrars further restricted discounts to new customers only. Although attackers may be sensitive to price restrictions, these limits might not deter them from acquiring many inexpensive domains, especially if they can automate account creation using an API, for instance.

\subsection{Payment Methods}

One important consideration for attackers is maintaining anonymity. While attackers might use stolen credit cards, we hypothesize that they may choose registrars that accept cryptocurrencies or digital wallets, as these add a layer of anonymity to the payment process. Out of 24 payment methods known to the TLD-List dataset, 13 are supported by the 31 analyzed registrars. Credit cards and PayPal stand out the most as they were available for 99.7\% and 98.9\% of domain registrations, respectively. Bitcoin goes next, supported in more than two-third of cases. Registrars proposed at least 1 but up to 7 different payment methods to choose from. Nevertheless, it is recognized that attackers often seek to conceal their identities when purchasing domain names. For example, while Porkbun accepts various forms of cryptocurrencies, they warn that the identity of the registrants may be verified so that they do not ``setup a phishing site, a fake store, or some other illegal or otherwise fraudulent/abusive site''~\cite{porkbun}.\footnote{Since the initial test, Porkbun has outsourced its crypto payments to Coinbase and no longer displays this disclaimer on its website.} Interestingly, only a small fraction of maliciously registered domains were purchased at Porkbun.

\begin{figure}[t]
    \centering
    \includegraphics[width=0.5\textwidth]{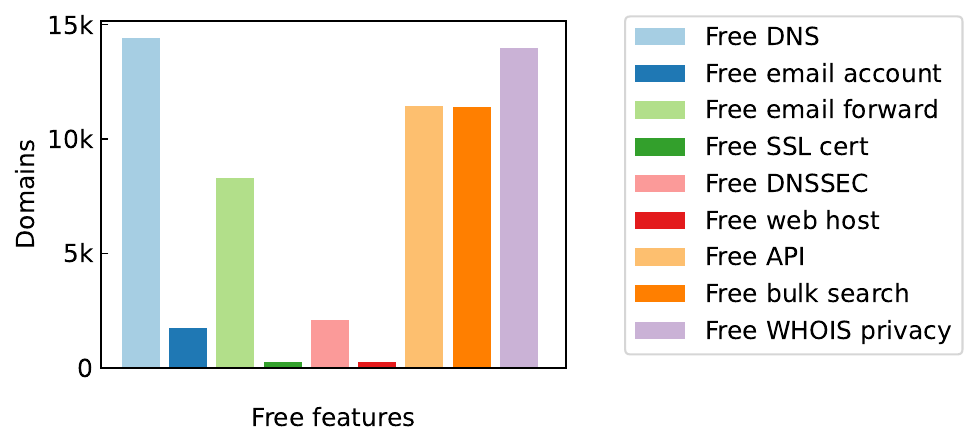}
    \caption{Nine free features proposed by registrars to their clients as well as the number of registrations that could have benefited from them.}
    \label{fig:free}
\end{figure}

\subsection{Free Features}

Whenever purchasing a domain name, registrants may receive various free services and add-ons. Figure~\ref{fig:free} shows the nine free features offered by registrars.

DNS service was offered the most, although not proposed by all the registrars for free. Free WHOIS privacy was available for the great majority of registrations (96.6\%) at 21 domain registrars. GoDaddy provides the ``Free Domain Privacy'' service for all the \textit{eligible} registrations but warns that some TLDs prohibit the use of WHOIS privacy services for its domains~\cite{godaddy}. Whenever WHOIS privacy is not provided for free, it is billed between \$0.18 and \$19.07 per year.

\begin{figure}[t]
    \centering
    \includegraphics[width=0.3\textwidth]{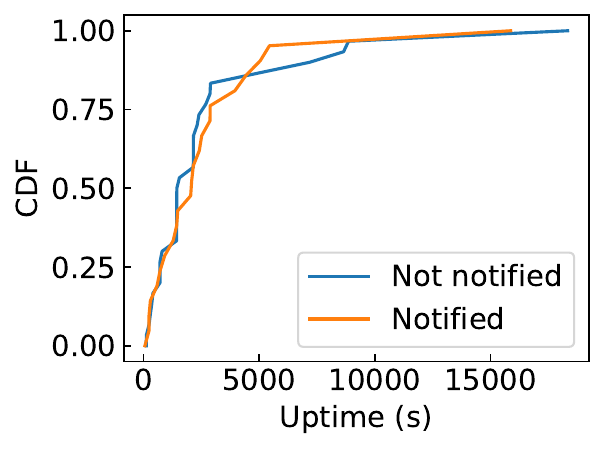}
    \caption{CDF of uptimes for notified and not notified registrars.}
    \label{fig:uptimes}
\end{figure}

A free API is offered by 16 of the 31 registrars we examined. We consider it free when provided without prerequisites, excluding cases in which it is only available to domain resellers. Four registrars permit the automated account creation, either as a new member of an existing organization or as a sub-account under an existing API user. Twelve registrars allow the automatic registration of domain names. Interestingly, the GoDaddy API was unrestricted during our analysis (August 2023 – January 2024), but as of June 2024, some features are limited to customers with a minimum number of domains or a Discount Domain Club subscription~\cite{godaddy-api}.

Attackers often register multiple domains in a single campaign~\cite{spamhaus-bulk,criminal-abuse,lifetimes}. We found that 18 registrars offer a ``bulk search'' feature, allowing clients to check availability and prices for multiple domains—ranging from 20, 5~K to 10~K+. Additionally, five registrars offer bulk registration discounts, with one registrar providing reduced prices for 10+ domains. Three registrars require a minimum number of managed domains for discounts, while two other offer them via sales inquiries.

Email forwarding, email account creation, SSL certificates, DNSSEC signing, and web hosting are offered less frequently. One registrar allows registrants to create up to 20 email aliases and forward emails to existing inboxes~\cite{square}. This offer does not include the creation of the email account itself, but 7 registrars out of 31 do provide such a service for free. Free web hosting and SSL certificates may facilitate setting up fully-fledged phishing websites but are available for only 254 and 235 registrations, respectively.

\subsection{Domain Name Registrant Data}

We next analyzed whether registrars \textit{proactively} verify that contact information is operational---ensuring email deliverability and phone number reachability. Registrants may be asked to provide personal information either when creating an account or during domain registration.

Our analysis shows that all registrars reject syntactically incorrect email addresses with a warning. To check verification, we used a valid email, and 26 out of 31 registrars sent confirmation emails. One registrar sends a password to ensure access to the mailbox, while two other registrars only send welcome emails without requiring action. Three registrars create accounts only at purchase, limiting email validation checks.

Phone number verification is much less common, with 23 registrars performing syntactic checks and only 6 conducting operational validation. For example, one registrar accepts syntactically incorrect phone numbers during registration but verifies them operationally when registering a domain, preventing attackers from using fake numbers. In contrast, another registrar requires registrants to choose between email and phone number verification before registering a new domain. 

Finally, we assess the syntactic validation of our physical address, once again not trivial to implement due to various formats imposed by different countries. Specifically, we supply a postal code without two last digits, in which case seven registrars signal the problem. For example, one registrar does the check based on the number of digits and warns that at least 4 are required. We also note that some registrars do not mark the postal code as a required field and therefore, they do not verify it. 

\subsection{Prevention of Trademark Infringement\label{sec:trademark}}

Whenever attackers target a specific brand, they want to ensure that the maliciously registered domain name looks as trustworthy as possible. For example, one may send a phishing email asking to visit the \texttt{https://my-paypal-account.com} web page to lure victims into providing their credentials. Registrars are well aware that certain customers may violate intellectual property rights and, therefore, provide the guidelines to solve copyright and trademark disputes. ICANN-accredited registrars refer to the ICANN Uniform Rapid Suspension System (URS)~\cite{urs} and the Uniform Domain-Name Dispute-Resolution Policy (UDRP)~\cite{udrp} for managing domain name disputes. 

However, these measures are \textit{reactive}, addressing issues only after a domain is registered and potentially used for malicious purposes. We assess whether registrars \textit{proactively} block attempts to purchase domains related to popular brands (\texttt{office365-my-account} and \texttt{facebook-login-page}) or those with random character sequences (\texttt{a9e86e6d5d4c676441da}). This evaluation covers every TLD/registrar pair in our dataset of 14.5~K maliciously registered phishing domains. 

Two registrars blocked trademarked domains from being added to the cart. One registrar shows an error: ``This domain contains restricted phrase(s) and can't be self-registered. Please contact support.'' Another registrar displays: ``We were unable to add the domain to the cart. Please contact support.'' Interestingly, an older account with the same registrar could still add branded domains, suggesting they might use reputation-based measures to prevent abuse by attackers creating multiple accounts. Interestingly, one registrar provides auto-generated suggestions for domain names. For instance, when attempting to add \texttt{facebook-login-page.company} to the cart, it suggests, ``This domain is suitable for a website that offers a secure and official login page for Facebook users.''

Lastly, when attempting to add a domain name containing a random string, none of the tested registrars triggered any error or warning. 

\subsection{Malicious Domain Uptimes}

Shorter uptimes should ideally discourage attackers from using certain TLDs and registrars, as swift suspension might drive them to seek alternatives. However, even brief activity may yield valuable credentials and financial gain, potentially diminishing the impact of reactive security measures on their registrar choices.

Malicious domains are often blocked by registrars after abuse reports or complaints. To investigate, we sampled domains from our dataset, submitted complaints, and compared their uptimes to those of unreported domains.

We notified 22 out of 31 registrars about 768 phishing domains. The uptimes of these domains varied up to 17 hours, with an average of 70 minutes. In comparison, domains not reported had a mean uptime of 61 minutes and a maximum of nearly 18 hours. We found little difference in uptimes between reported and unreported domains.

To gain deeper insight, we analyze the uptimes of reported and non-reported domains aggregated at the registrar level instead of at the domain level as shown in Figure~\ref{fig:uptimes}. Among the notified registrars, three of them had short median uptimes, taking just a few minutes each. Conversely, the domains registered with three other registrars exhibited relatively longer median uptimes, at 84 minutes, 1.5 hours, and nearly 4.5 hours.

Similar results at the domain and registrar levels may be due to registrars subscribing to reputable abuse feeds, enabling quick mitigation of phishing domains, or to concurrent notification campaigns.

\section{Driving Factors of Domain Abuse\label{sec-driving-factors}}

In the previous sections, we have identified various registration attributes and practices that may influence the attacker preference when maliciously registering a domain name. In this section, we develop two models to estimate and statistically demonstrate which features have an important impact and quantify its magnitude.

\subsection{Feature Engineering \label{sec:feature-eng}}

Given the high dimensionality of the initial feature set, it was essential to undergo a feature engineering phase to refine the model by selecting the most relevant features. On a first phase of the feature engineering process, we merged features that represent similar underlying constructs. For example, the features related to digital wallets (\textit{payment_alipay}, \textit{payment_applepay}, etc.) all represent different forms of digital payment methods. By aggregating these into a single binary indicator (\textit{payment_digital_wallet}), we effectively capture the broader concept of ``digital payment method availability'' rather than treating each form of digital wallet as an independent predictor. This approach reduces multicollinearity, as similar variables can inflate the variance of coefficient estimates, leading to less reliable models. Similarly, grouping payment methods into categories such as \textit{payment_crypto} and \textit{payment_transfer} consolidates the model to focus on the higher-level types of payment methods rather than individual options. This aggregation maintains the interpretability of the feature and aligns with the idea that different payment methods within a category are likely to have similar effects on the dependent variable.

Additionally, following the same rational, we also aggregated prevention measures (\textit{prevention}), restrictions (all the boolean \textit{restriction_*} features), personal data validation (\textit{emailPhone_validated}), and API offerings (\textit{API}). These features are likely to have correlated effects on domain abuse, and summing them into composite indicators captures the overall presence or absence of these protective measures rather than assessing them separately. This not only simplifies the model but also aligns with the principle of parsimony in statistical modeling, where the goal is to explain the data with the fewest possible predictors.

As for the computation of the average maliciously registered domain uptime, combining the uptime of notified and not notified domains by averaging  produces a single representative measure of uptime for each registrar. Uptime, as a feature (\textit{uptime}), theoretically could influence the likelihood of malicious registrations, and having a unified metric simplifies the model without losing relevant information. 

Next, to create a parsimonious model, we selected features, ensuring that the model focuses on features with the highest impact, which might not be immediately apparent from a purely statistical or automated feature selection process. 
We identified the following categories of features as the most relevant to malicious domain registrations: 
\textit{free\_bulk\_search\_number}, \textit{price\_register}, \textit{discount\_register}, \textit{restrictions}, \textit{prevention}, \textit{API}, \textit{payment\_digital\_wallet}, \textit{payment\_crypto}, \textit{payment\_transfer}, \textit{uptime}, \textit{emailPhone\_validated}, \textit{free\_dns}, \textit{free\_web\_host}, and \textit{free\_ssl\_cert}. These features span several important dimensions.

Pricing and discount features, like \textit{price\_register} and \textit{discount\_register}, were selected based on the hypothesis that lower costs might attract a higher number of abusive registrations, as malicious entities typically operate with limited budgets. Registrar restrictions and prevention measures, including \textit{restrictions} and \textit{prevention}, were chosen to capture the extent to which registrars enforce policies that could mitigate domain abuse. Technical and payment features, such as \textit{API}, \textit{payment\_digital\_wallet}, \textit{payment\_crypto}, and \textit{payment\_transfer}, reflect the technical and financial infrastructure that can either facilitate or hinder domain abuse. The operational feature \textit{emailPhone_validated} was included to assess the operational reliability of identity validation procedures, which may play a role in preventing the registration of malicious domains. Finally, free services, including variables such as \textit{free\_dns}, \textit{free\_web\_host}, and \textit{free\_ssl\_cert}, were chosen because offering free services can lower the barriers to entry for malicious actors who seek to register domains at minimal cost.

\subsection{Model\texorpdfstring{$_1$}\mbox{:} GLM Negative Binomial Regression}

\begin{figure}[t]
    \centering
    \includegraphics[width=0.8\columnwidth]{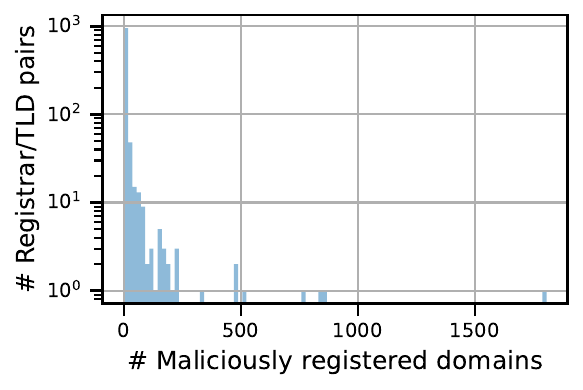}
    \caption{Distribution of maliciously registered domain names per registrar/TLD pair.}
    \label{fig:malicioushist}
\end{figure}

We used a Generalized Linear Model (GLM) with negative binomial regression to estimate the impact of features on malicious domain counts per registrar/TLD pair. This model handles overdispersed count data, where variance exceeds the mean, typical in domain abuse distributions (see Figure~\ref{fig:malicioushist}).

This model is well-suited to our analysis due to its ability to handle high variance in domain abuse counts, often caused by ``super-spreader'' registrars or TLDs (see Table~\ref{registrars-tlds-malicious}). Unlike the Poisson model, which assumes equal mean and variance, the negative binomial model addresses overdispersion and provides more reliable estimates for this specific count data where there is a lot of variance. Additionally, the GLM framework offers clear coefficient interpretation.

\begin{figure}[t]
    \centering
    \includegraphics[width=\columnwidth]{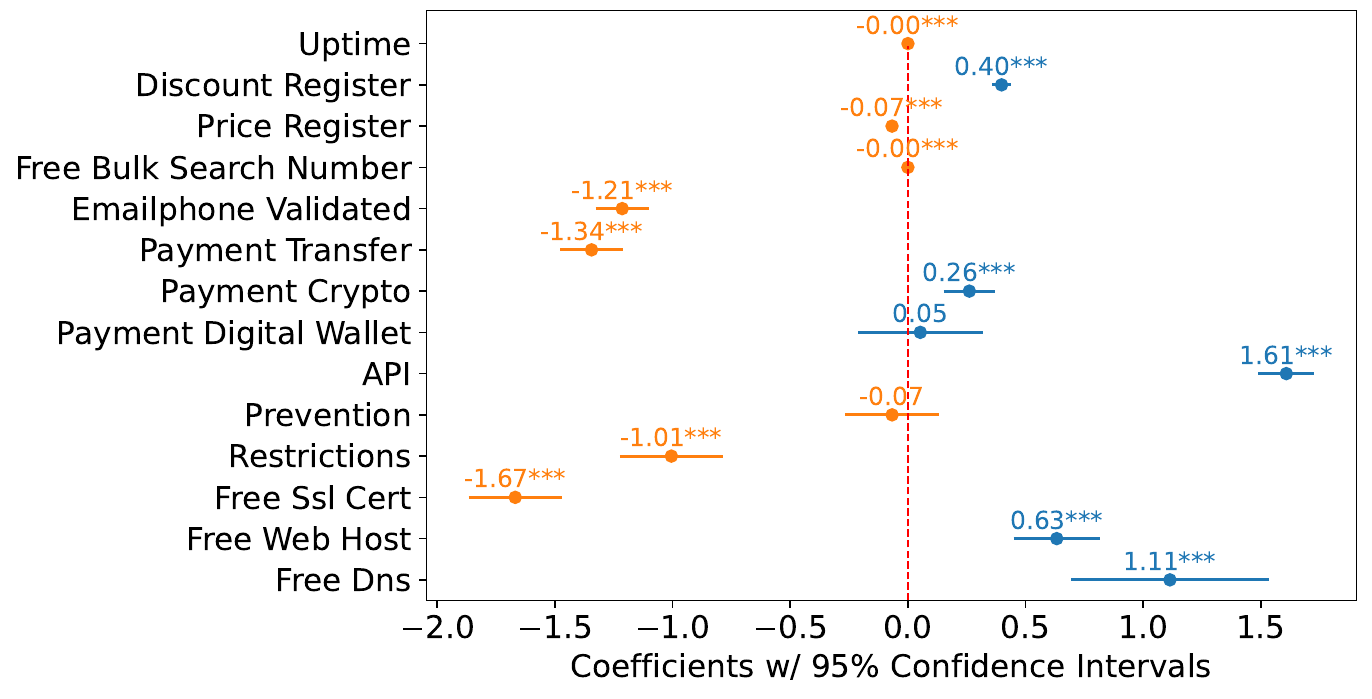}
    \caption{GLM model: estimated coefficients with the 95\% confidence intervals.}
    \label{fig:coeffs}
\end{figure}

\textbf{Model$_1$ Results:} Table~\ref{tab:coeffs} in Appendix shows the results after estimating the model. It included 1,066 observations with 14 features and a constant term. It achieved a pseudo R-squared value of 0.7733 indicating that the model explains approximately 77.33\% of the variance in the number of malicious domains per registrar/TLD pair. Figure~\ref{fig:coeffs} shows the summary of the results. Exponentiating the coefficients of the fitted model allows us to interpret them as a multiplicative factors for the dependent variable, the number of maliciously registered domains in this case. In particular, several \textit{registration attributes} have a statistically significant effect:

\begin{itemize}
    \item Registration price has a coefficient of -0.07 (p $<$ 0.001). Exponentiating this coefficient gives \(e^{-0.07} \approx 0.94\), indicating that decreasing the registration price by one dollar is associated with a 6.6\% increase in the number of malicious domains, suggesting that more affordable registration fees may encourage higher rates of abuse.
    
    \item Registration discounts have a positive coefficient of 0.40 (p~$<$~0.001), which suggests that offering a one dollar discount on domain registration is associated with a 49\% increase in malicious registrations, highlighting a potential incentive for malicious actors to exploit discounts.
       
    \item Cryptocurrency payments show a positive coefficient of 0.26 (p = 0.017), which implies a 30\% increase in malicious registrations when cryptocurrency payments are accepted. Conversely, transfer-like payments have a negative coefficient of -1.34 (p~$<$~0.001), suggesting a 74\% decrease in malicious domains with the acceptance of bank transfers.

    \item On the technical side, the presence of APIs either to register domains or to create accounts have a positive coefficient of 1.60 (p~$<$~0.001), which indicates that registrars offering API access are associated with a 401\% increase in the number of malicious domains.   
    
    \item Free services have a positive coefficient of 1.11 (p~=~0.008), which means that they are associated with approximately a 205\% increase in the number of maliciously registered domains compared to registrars without them.  Similarly, the availability of free web hosting shows a positive coefficient of 0.63 (p~=~0.001) indicating that free web hosting is associated with an 88\% increase in the number of malicious phishing domains. In contrast, offering free SSL certificates  has a negative coefficient of -1.67 (p~$<$~0.001) meaning that it is associated with an 81\% decrease in the number of malicious registrations.
 
\end{itemize}

Focusing on the \textit{proactive} verification, the restrictions implemented by some registrars have a negative coefficient of -1.01 (p $<$ 0.001), which  suggests that stringent registrar restrictions are associated with a 63\% decrease in the number of maliciously registered domains. Similarly, when the validation of registrant information such as their phone number of email takes place during the account creation or before the domain purchase, it has a significant negative coefficient of -1.21 (p $<$ 0.001) indicating that it is associated with a 70\% decrease in malicious registrations.

When it comes to reactive security practices, \textit{uptime} has a small coefficient of -0.0001, which indicates that higher uptimes are weakly associated  with a very slight decrease in the number of malicious registrations suggesting that it has a minor impact and its effect on reducing domain abuse is relatively small.

\subsection{Model\texorpdfstring{$_2$}\mbox{:} Multilevel Logistic Regression}

The second model is designed to quantify the impact of the identified features on the probability of a domain being registered with a specific registrar for malicious or legitimate  purposes. This model uses a multilevel hierarchical logistic regression approach, well suited for handling the nested structure of the data, in which domains are clustered within registrars and TLDs. By accounting for this hierarchical structure, the model can more accurately estimate the impact of registrar-specific and TLD-specific practices and attributes on the likelihood of domain abuse.

In this model, the dependent variable is defined as the binary status of a domain, where \texttt{True} indicates that the domain was registered with malicious intent, and \texttt{False} signifies that the domain was registered for legitimate purposes. The independent variables, which include the  features identified as potentially influencing domain abuse, are modeled as fixed effects. They capture the direct impact of each feature on the probability of a domain being maliciously registered.

The hierarchical structure of the model is captured by including two levels of random effects: one for the registrar and another for the TLD. The registrar-level random effect allows the model to account for variability between registrars that might not be fully explained by the fixed effects such as differences in registrar practices or market strategies. Similarly, the TLD-level random effect accounts for the variability between different TLDs recognizing that the domains within the same TLD might exhibit similar patterns of abuse due to the factors specific to that TLD. By incorporating both registrar-level and TLD-level random effects, the model adjusts for the within-group correlations at each level providing more reliable estimates of the fixed effects. 

\textbf{Model$_2$ Results:} Table~\ref{tab:rcoeffs} in Appendix shows the results after estimating the model.  The conditional $R^2$ for the full model, which incorporates both fixed and random effects, is 0.47, which means that approximately 47.4\% of the variance in the probability of domains being registered for malicious purposes is explained when considering the complete structure of the model, including the effects at both the registrar and TLD levels. On the other hand, the marginal $R^2$, which reflects the proportion of variance explained solely by the fixed effects, is considerably lower at 0.11. This difference underscores the significant contribution of the random effects to the model explanatory power, indicating that a substantial portion of the variability in domain abuse is due to the differences at the registrar and TLD levels, beyond what can be captured by the fixed effects alone.

\begin{figure}[t]
    \centering
    \includegraphics[width=0.8\columnwidth]{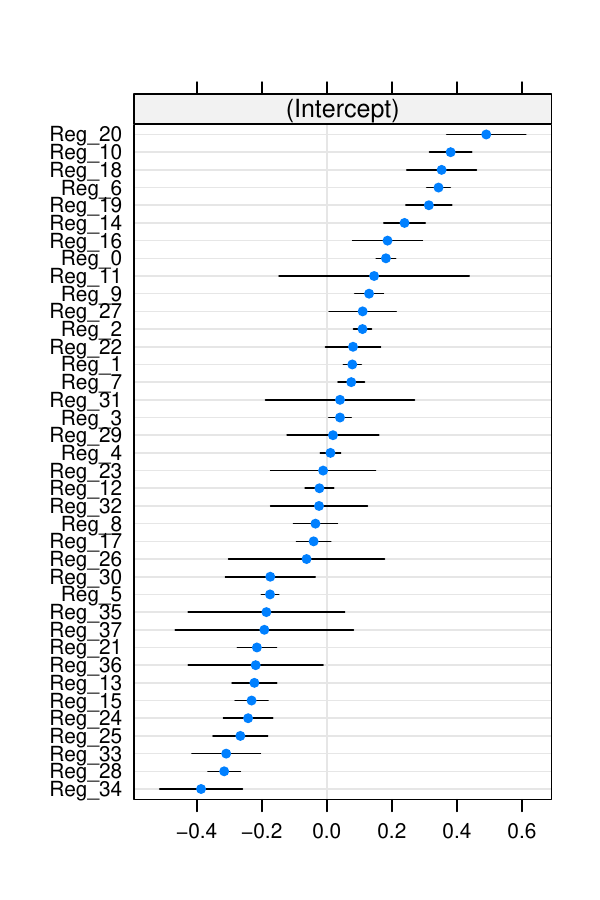}
    \caption{Random effects at the registrar level derived from the second model.}
    \label{fig:regrandom}
\end{figure}

When examining the random effects at the registrar level, the model reveals a variance of 0.06 with a standard deviation of 0.26. Figure~\ref{fig:regrandom} shows these estimates. For example, Reg\_20 has a random intercept of 0.49, which indicates that this registrar has a higher likelihood of having malicious domain registrations compared to the average registrar. Specifically, this positive value suggests that domains registered through Reg\_20 are more likely to be malicious than those registered through registrars with lower or negative intercepts.  Conversely, Reg\_34 has a random intercept of -0.38, indicating that domains registered through this registrar are less likely to be malicious compared to the average. This negative value implies that the Reg\_34 practices or characteristics are associated with a lower probability of domain abuse, making it a less attractive option for malicious registrants. These intercepts highlight how individual registrar characteristics significantly impact the likelihood of domain abuse.

At the TLD level, the variance of the random intercept is estimated at 0.03, with a standard deviation of 0.16 (see Figure \ref{fig:figureTLDs} in  Appendix). Although there is some variability in domain abuse likelihood across different TLDs, this variance is modest compared to that observed at the registrar level. The smaller variance at the TLD level indicates that while TLD-specific characteristics do influence domain abuse, their impact is less pronounced than that of registrar-specific factors, which is expected. 

\begin{figure}[t]
    \centering
    \includegraphics[width=\columnwidth]{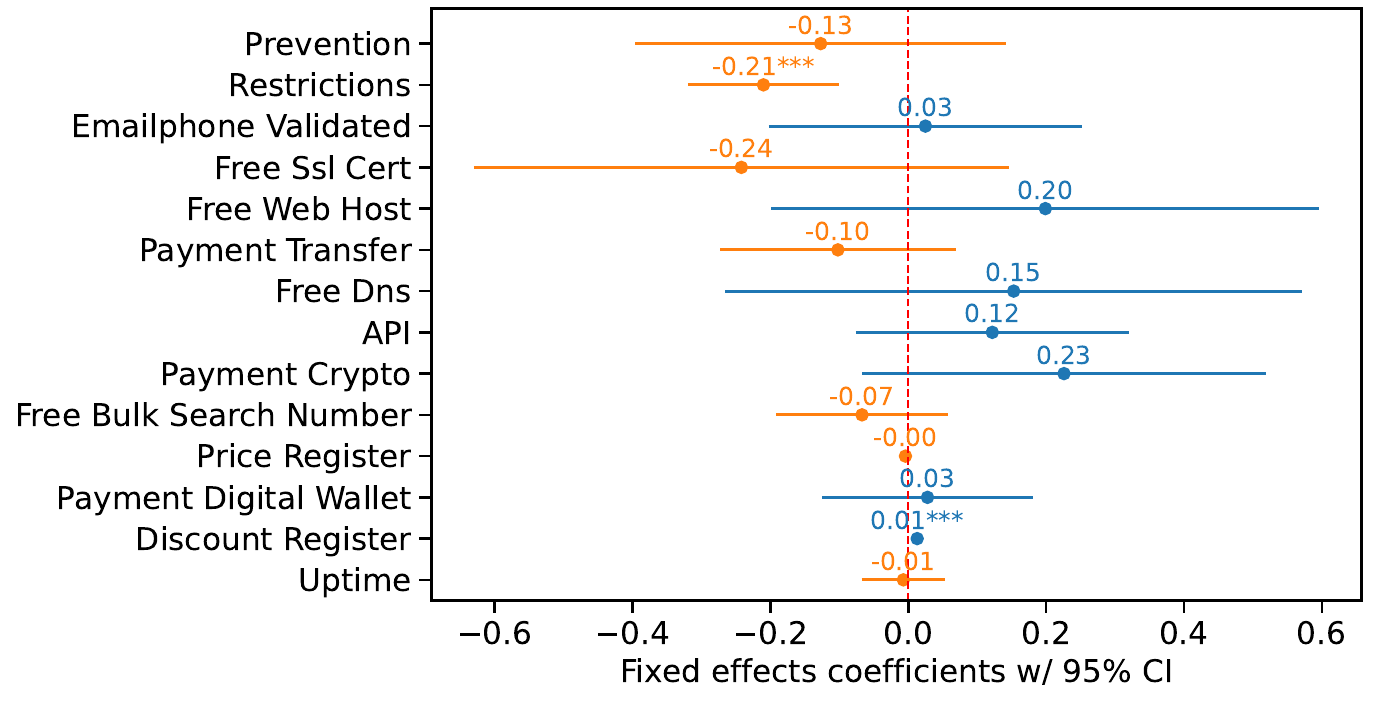}
    \caption{Fixed effect coefficients with 95\% confidence intervals derived from the second model.}
    \label{fig:fecoeff}
\end{figure}

On the other hand, the fixed effects in the model reveal key insights into how certain features influence the likelihood of a domain being registered with malicious intent. Notably, registration discounts and restrictions emerged as significant features (see Figure~\ref{fig:fecoeff}). Registration discounts have a positive coefficient (0.013, p~$<$~0.001), indicating that the domains registered with discounts are more likely to be malicious, which means that for every unit increase in the discount, the odds of a domain being maliciously registered increase by about 1.3\%. For instance, if a domain registration price is reduced by \$10, the odds of that domain being maliciously registered would increase by approximately 13.8\%.  These values suggest that attackers may be drawn to registrars offering discounts, possibly due to lower costs facilitating large-scale domain abuse. When considering legitimate registrations, the significance of discounts in attracting malicious registrations implies that promotions may be less critical for legitimate users. While discounts appear to be a strong motivator for malicious actors—likely because they reduce the financial barrier for bulk domain registrations used in various forms of online abuse—legitimate registrants might prioritize other factors over cost savings. 

Conversely, restrictions show a negative coefficient (-0.210, p~$<$~0.001), implying that registrars with stringent registration restrictions are associated with a reduced likelihood of malicious domain registrations, which means that the presence of restrictions decreases the odds of a domain being maliciously registered by about 19\%. While it is an effective deterrent for malicious actors, it could also imply that legitimate registrants are subjected to a more rigorous registration process, which may deter those seeking a quicker or less cumbersome registration experience.

Other variables such as registration price, average domain uptime, and payment types were not statistically significant, suggesting they do not have a consistent impact on the likelihood of a domain being registered for malicious purposes. This indicates that changes in these features do not clearly distinguish between malicious and benign registrations, and their influence appears similar for both.
\section{Discussion \label{sec:discussion}}

This section synthesizes the findings from our analysis. While the first model identifies the factors driving the likelihood of DNS abuse concentrations for policy differences between TLD-registrar pairs, the second one indicates whether the registrar  or TLD level features are favored by attackers alone or also by legitimate users. Our exploration of the phishing domain landscape highlights the strategic choices of phishers within the domain registration ecosystem. 

\textbf{Impact of economic incentives:}
Both models consistently show that economic incentives, such as registration discounts, are associated with an increase in the number of malicious registrations. The statistical analysis supports the descriptive findings revealing that nearly half of the malicious domains cost \$2 or less. The presence of discounts raises the likelihood of domains being registered for malicious purposes. Even if discounts are limited to new users, attackers may exploit free APIs to automate account creation and domain registration at discounted prices. By leveraging low-cost options, attackers can maximize their return on investment, especially given the short lifespan of these domains before they are suspended.

\textbf{Role of free services:} The first model demonstrates that free services (web hosting and DNS) significantly increases the number of malicious registrations, indicating that they lower entry barriers for attackers, allowing them to set up and maintain malicious domains with minimal expense. However, the results of the second model  suggest that they are attractive to both malicious and legitimate users, which is unsurprising.   

More importantly, the availability of APIs for domain registration and account management is strongly associated with a higher volume malicious registrations. APIs simplify can automate the account and domain registration process, potentially enabling rapid, large-scale abuse and attracting attackers who seek to exploit registration procedures and policies efficiently. 

\textbf{Payment methods offerings:} Cryptocurrency payments are linked to higher volumes of malicious registrations, while transfer-like payments are associated with lower volumes, which suggests that some payment methods facilitate malicious registrations, while others act as deterrents due to their traceability and additional verification requirements. However, no significant differences were observed in payment methods between attackers and legitimate users, indicating that these options serve both types of registrations. 

\textbf{Proactive restrictions:} The implementation of proactive restrictions, such as stringent registration policies and mandatory email/phone validation, was associated with a significant decrease in malicious domain registrations, which demonstrates the effectiveness of upfront checks and restrictions in preventing abuse before it occurs. While attackers can acquire disposable emails and temporary phone numbers, the results highlight the effectiveness of upfront checks and restrictions in preventing abuse. 

\textbf{Reactive measures:} While promptly suspending malicious domain names is essential for mitigating potential harm, our analysis shows that longer uptime has only a marginal effect on the concentration of malicious domains and minimal impact on the attacker choice of a registrar or TLD. Even short operational periods can provide attackers with valuable credentials and financial rewards, making reactive security measures focused on uptime of limited influence.

Finally, our analysis suggests that while initial pricing plays a role, attractiveness to attackers likely results from a combination of factors---for example, purchasing higher-cost domains may still be appealing if discounts are available or payment in Bitcoin is accepted. 

\section{Conclusions\label{sec:conclusions}}

This study sheds light on the factors that drive malicious domain registrations, particularly in phishing campaigns. We have identified key trends showing that attackers exploit cost-effective registration processes and automation features to facilitate their activities. Specifically, price reductions, free services, and API access significantly increase the likelihood of domain abuse. 

Our findings emphasize that attackers prioritize registrars offering lower costs and features that enable bulk registration, which allows them to scale their operations efficiently. In contrast, stringent registration restrictions serve as effective deterrents, reducing the probability of malicious domains being registered. 

This research provides a clearer understanding of the strategies employed by cybercriminals when selecting registrars and top-level domains. By focusing on the features that attackers most commonly exploit, it becomes possible to design more targeted interventions to disrupt their operations while still supporting legitimate domain registrations. 

\section*{Acknowledgments}
We would like to thank the Anti-Phishing Working Group, Phishtank, OpenPhish for providing data access for this analysis and to Spamhaus and SURBL for granting access to their blocklists, which helped us filter out malicious domains from the benign list we sampled. 

\bibliographystyle{plain}
\bibliography{reference}

\section*{Appendix}

\begin{table}[h]
    \caption{Distribution of maliciously registered domain names across 20 registrars responsible for 83.8\% of registrations. Overall, domain names originated from 157 registrars.} 
    \label{registrars-domains-all} 
    \scriptsize
     \setlength{\tabcolsep}{4pt}
    \centering 
    \begin{tabular}{cclc}
        \toprule
        \textbf{Rank} & \textbf{IANA ID} & \textbf{Registrar Name} & \textbf{Domains}\\
        \midrule
            1. & 1479 & NameSilo, LLC & 5,986 \\
            2. & 1636 & HOSTINGER operations, UAB & 2,558 \\
            3. & 1923 & Gname.com Pte. Ltd. & 1,958 \\
            4. & 3765 & NICENIC INTERNATIONAL GROUP CO [...] & 1,932 \\
            5. & 1068 & NameCheap, Inc. & 1,711 \\
            6. & 146 & GoDaddy.com, LLC & 1,383 \\
            7. & 1606 & Registrar of Domain Names REG.RU LLC & 1,093 \\
            8. & 303 & PDR Ltd. d/b/a PublicDomainRegistry.com & 1,061 \\
            9. & 3775 & ALIBABA.COM SINGAPORE [...] & 763 \\
            10. & 895 & Squarespace Domains II LLC & 751 \\
            11. & 3858 & Aceville Pte. Ltd. & 675 \\
            12. & 69 & Tucows Domains Inc. & 654 \\
            13. & 609 & Sav.com, LLC & 621 \\
            14. & 1250 & OwnRegistrar, Inc. & 566 \\
            15. & 472 & Dynadot Inc & 487 \\
            16. & 2 & Network Solutions, LLC & 327 \\
            17. & 1555 & 22net, Inc. & 311 \\
            18. & 1599 & Alibaba Cloud Computing Ltd. d/b/a HiChina [...] & 282 \\
            19. & 49 & GMO Internet Group, Inc. d/b/a Onamae.com & 246 \\
            20. & 625 & Name.com, Inc. & 242 \\
        \bottomrule
    \end{tabular}
\end{table}

\begin{table}[t]
    \caption{Top 20 most common top-level domains in the list of 14,474 maliciously registered domains with the TLD-List data.} 
    \label{tlds} 
    \scriptsize
    \centering 
    \begin{tabular}{cccc}
        \toprule
        \textbf{Rank} & \textbf{TLD} & \textbf{Type} & \textbf{Count}\\
        \midrule
            1. & \texttt{com} & generic & 3,850 \\
            2. & \texttt{top} & generic & 2,002 \\
            3. & \texttt{online} & generic & 1,179 \\
            4. & \texttt{shop} & generic & 858 \\
            5. & \texttt{info} & generic & 807 \\
            6. & \texttt{xyz} & generic & 699 \\
            7. & \texttt{site} & generic & 419 \\
            8. & \texttt{cloud} & generic & 341 \\
            9. & \texttt{buzz} & generic & 276 \\
            10. & \texttt{us} & country-code & 260 \\
            11. & \texttt{net} & generic & 242 \\
            12. & \texttt{org} & generic & 236 \\
            13. & \texttt{sbs} & generic & 231 \\
            14. & \texttt{click} & generic & 216 \\
            15. & \texttt{life} & generic & 191 \\
            16. & \texttt{cc} & country-code & 177 \\
            17. & \texttt{live} & generic & 158 \\
            18. & \texttt{pro} & generic-restricted & 145 \\
            19. & \texttt{cfd} & generic & 144 \\
            20. & \texttt{icu} & generic & 101 \\
        \bottomrule
    \end{tabular}
\end{table}

\begin{table}[t]
    \caption{All the registrars that registered malicious domain names in our dataset.} 
    \label{registrars-malicious} 
    \scriptsize
    \centering 
    \begin{tabular}{clcc}
        \toprule
        \textbf{Rank} & \textbf{Registrar} & \textbf{IANA ID(s)} & \textbf{Domains}\\
        \midrule
            1. & NameSilo & 1479 & 5,751 \\
            2. & Hostinger & 1636 & 2,473 \\
            3. & Namecheap & 1068 & 1,649 \\
            4. & GoDaddy & 146 & 1,328 \\
            5. & Alibaba Cloud & 3775, 1599 & 897 \\
            6. & Sav & 609, 3892, 3893, 3895 & 603 \\
            7. & Dynadot & 472 & 471 \\
            8. & Name.com & 625 & 228 \\
            9. & Cloudflare Registrar & 1910 & 203 \\
            10. & Porkbun & 1861 & 167 \\
            11. & INWX & 1420 & 159 \\
            12. & Cosmotown & 1509 & 103 \\
            13. & Regtons.com & 1505 & 80 \\
            14. & Amazon Route 53 & 468 & 53 \\
            15. & Spaceship & 3862 & 49 \\
            16. & DreamHost & 431 & 31 \\
            17. & Hello & 1924 & 31 \\
            18. & Internet.bs & 2487 & 29 \\
            19. & Domain.com & 886 & 27 \\
            20. & OVH & 433 & 23 \\
            21. & Netim & 1519 & 23 \\
            22. & Above.com & 940 & 22 \\
            23. & BigRock & 1495 & 21 \\
            24. & Gandi & 81 & 19 \\
            25. & 123 Reg & 1515 & 9 \\
            26. & 101domain & 1011 & 9 \\
            27. & Dreamscape Networks & 1291 & 5 \\
            28. & Instra Corporation & 1376 & 4 \\
            29. & EuroDNS & 1052 & 3 \\
            30. & Reg.com & 1606 & 2 \\
            31. & alldomains.hosting & 809 & 2 \\
        \bottomrule
    \end{tabular}
\end{table}

\begin{table}[t]
    \scriptsize
    \centering
    \caption{All collected registration features  and indication whether they are used at the modeling stage.}\label{tab:features-registration}
    \begin{tabular}{llcc}
        \toprule
        \textbf{\#} & \textbf{Feature Name} & \textbf{Feature Type} & \textbf{Source} \\
        \midrule
            1. & \textit{free\_api} & Boolean & Manual \\
            2. & \textit{api\_create\_account} & Boolean & Manual \\
            3. & \textit{api\_register\_domain} & Boolean & Manual \\
            4. & \textit{free\_dns} & Boolean & TLD-List \\
            5. & \textit{free\_dnssec} & Boolean & Manual \\
            6. & \textit{free\_email\_account} & Boolean & TLD-List \\
            7. & \textit{free\_email\_forward} & Boolean & TLD-List \\
            8. & \textit{free\_web\_hosting} & Boolean & Manual \\
            9. & \textit{free\_ssl\_cert} & Boolean & TLD-List \\
            10. & \textit{free\_bulk\_search\_number} & Numerical & Manual \\
            11. & \textit{bulk\_discount} & Boolean & Manual \\
            12. & \textit{payment\_alipay} & Boolean & TLD-List \\
            13. & \textit{payment\_applepay} & Boolean & TLD-List \\
            14. & \textit{payment\_banktransfer} & Boolean & TLD-List \\
            15. & \textit{payment\_bitcoin} & Boolean & TLD-List \\
            16. & \textit{payment\_cashinperson} & Boolean & TLD-List \\
            17. & \textit{payment\_cc} & Boolean & TLD-List \\
            18. & \textit{payment\_check} & Boolean & TLD-List \\
            19. & \textit{payment\_dinersclub} & Boolean & TLD-List \\
            20. & \textit{payment\_dwolla} & Boolean & TLD-List \\
            21. & \textit{payment\_giropay} & Boolean & TLD-List \\
            22. & \textit{payment\_googlewallet} & Boolean & TLD-List \\
            23. & \textit{payment\_moneyorder} & Boolean & TLD-List \\
            24. & \textit{payment\_neteller} & Boolean & TLD-List \\
            25. & \textit{payment\_payeer} & Boolean & TLD-List \\
            26. & \textit{payment\_paypal} & Boolean & TLD-List \\
            27. & \textit{payment\_payza} & Boolean & TLD-List \\
            28. & \textit{payment\_qiwi} & Boolean & TLD-List \\
            29. & \textit{payment\_skril} & Boolean & TLD-List \\
            30. & \textit{payment\_topcoin} & Boolean & TLD-List \\
            31. & \textit{payment\_webmoney} & Boolean & TLD-List \\
            32. & \textit{payment\_westernunion} & Boolean & TLD-List \\
            33. & \textit{payment\_worldpay} & Boolean & TLD-List \\
            34. & \textit{payment\_yandexmoney} & Boolean & TLD-List \\
            35. & \textit{payment\_yoomoney} & Boolean & TLD-List \\
            36. & \textit{price\_register} & Numerical & TLD-List \\
            37. & \textit{price\_renewal} & Numerical & TLD-List \\
            38. & \textit{price\_transfer} & Numerical & TLD-List \\
            39. & \textit{price\_whois\_privacy} & Numerical & TLD-List \\
            40. & \textit{discount\_register} & Numerical & TLD-List \\
            41. & \textit{discount\_renewal} & Numerical & TLD-List \\
            42. & \textit{discount\_transfer} & Numerical & TLD-List \\
            43. & \textit{term\_new\_customer\_only\_register} & Boolean & TLD-List \\
            44. & \textit{term\_new\_customer\_only\_transfer} & Boolean & TLD-List \\
            45. & \textit{term\_limit\_per\_customer\_register} & Numerical & TLD-List \\
            46. & \textit{term\_limit\_per\_customer\_transfer} & Numerical & TLD-List \\
        \bottomrule
    \end{tabular}
\end{table}

\begin{table}[t]
    \scriptsize
    \centering
    \caption{All collected proactive/reactive features  and indication whether they are used at the modeling stage.}\label{tab:features-proactive-reactive}
    \begin{tabular}{llcc}
        \toprule
        \textbf{\#} & \textbf{Feature Name} & \textbf{Feature Type} & \textbf{Source} \\
        \midrule
            1. & \textit{email\_syntactically\_validated} & Boolean & Manual \\
            2. & \textit{phone\_syntactically\_validated} & Boolean & Manual \\
            3. & \textit{address\_syntactically\_validated} & Boolean & Manual \\
            4. & \textit{email\_operational\_validated} & Boolean & Manual \\
            5. & \textit{phone\_operational\_validated } & Boolean & Manual \\
            6. & \textit{random\_warning} & Boolean & Manual \\
            7. & \textit{random\_prevention} & Boolean & Manual \\
            8. & \textit{office365\_warning} & Boolean & Manual \\
            9. & \textit{office365\_prevention} & Boolean & Manual \\
            10. & \textit{facebook\_warning} & Boolean & Manual \\
            11. & \textit{facebook\_prevention} & Boolean & Manual \\
            12. & \textit{restriction\_not\_available} & Boolean & Manual \\
            13. & \textit{restriction\_local\_presence} & Boolean & Manual \\
            14. & \textit{restriction\_community\_ties} & Boolean & Manual \\
            15. & \textit{restriction\_age\_restriction} & Boolean & Manual \\
            16. & \textit{restriction\_infrastructure} & Boolean & Manual \\
            17. & \textit{restriction\_group\_ties} & Boolean & Manual \\
            18. & \textit{restriction\_commitment\_required} & Boolean & Manual \\
            19. & \textit{restriction\_id\_required} & Boolean & Manual \\
            20. & \textit{restriction\_region\_ties} & Boolean & Manual \\
            21. & \textit{restriction\_professionals\_only} & Boolean & Manual \\
            22. & \textit{restriction\_certain\_nationals\_prohibited} & Boolean & Manual \\
            23. & \textit{restriction\_org\_or\_affiliates\_only} & Boolean & Manual \\
            24. & \textit{restriction\_exclusive\_registrar} & Boolean & Manual \\
            25. & \textit{restriction\_content\_restrictions} & Boolean & Manual \\
        \midrule
            26. & \textit{uptime\_notified} & Numerical & Automated \\
            27. & \textit{uptime\_not\_notified} & Numerical & Automated \\
        \bottomrule
    \end{tabular}
\end{table}

\begin{table}[t]
\scriptsize
\begin{center}
    \caption{Generalized Linear Model Regression Results}\label{tab:coeffs}
\begin{tabular}{lclc}

\toprule
\textbf{Dep. Variable:}                   &    malicious     & \textbf{  No. Observations:  } &     1066    \\
\textbf{Model:}                           &       GLM        & \textbf{  Df Residuals:      } &     1051    \\
\textbf{Model Family:}                    & NegativeBinomial & \textbf{  Df Model:          } &       14    \\
\textbf{Link Function:}                   &       Log        & \textbf{  Scale:             } &    1.0000   \\
\textbf{Method:}                          &       IRLS       & \textbf{  Log-Likelihood:    } &   -3093.9   \\
\textbf{Pearson chi2:      } &  1.07e+04  & \textbf{  Deviance:          } &    2970.6   \\
\textbf{No. Iterations:}                  &        65        & \textbf{  Pseudo R-squ. (CS):} &   0.7733    \\
\bottomrule
\end{tabular}
\scalebox{.8}{
\begin{tabular}{lcccccc}
                                          & \textbf{Coef} & \textbf{std err} & \textbf{z} & \textbf{P$>|z|$} & \textbf{[0.025} & \textbf{0.975]}  \\
\midrule
\textbf{Intercept}                        &       2.3927  &        0.355     &     6.748  &         0.000        &        1.698    &        3.088     \\
\textbf{Free DNS}                &       1.1134  &        0.420     &     2.648  &         0.008        &        0.289    &        1.937     \\
\textbf{Free Web host}          &       0.6323  &        0.183     &     3.458  &         0.001        &        0.274    &        0.991     \\
\textbf{Free SSL cert}          &      -1.6688  &        0.198     &    -8.440  &         0.000        &       -2.056    &       -1.281     \\
\textbf{Restrictions}             &      -1.0053  &        0.219     &    -4.594  &         0.000        &       -1.434    &       -0.576     \\
\textbf{Prevention}               &      -0.0673  &        0.200     &    -0.336  &         0.737        &       -0.460    &        0.325     \\
\textbf{API}                      &       1.6080  &        0.118     &    13.585  &         0.000        &        1.376    &        1.840     \\
\textbf{Payment digital wallet} &       0.0525  &        0.264     &     0.199  &         0.843        &       -0.466    &        0.571     \\
\textbf{Payment crypto}          &       0.2609  &        0.109     &     2.393  &         0.017        &        0.047    &        0.475     \\
\textbf{Payment transfer}        &      -1.3446  &        0.133     &   -10.131  &         0.000        &       -1.605    &       -1.084     \\
\textbf{EmailPhone validated}    &      -1.2143  &        0.113     &   -10.757  &         0.000        &       -1.436    &       -0.993     \\
\textbf{Free bulk search}       &      -0.0003  &     5.38e-05     &    -5.687  &         0.000        &       -0.000    &       -0.000     \\
\textbf{Price register}                  &      -0.0676  &        0.005     &   -14.565  &         0.000        &       -0.077    &       -0.058     \\
\textbf{Discount register}               &       0.3979  &        0.040     &     9.957  &         0.000        &        0.320    &        0.476     \\
\textbf{Uptime}                           &      -0.0001  &     2.54e-05     &    -4.660  &         0.000        &       -0.000    &    -6.85e-05     \\
\bottomrule
\end{tabular}}

\end{center}
\end{table}

\begin{table}[ht]
\scriptsize
\centering
    \caption{Multilevel Logistic Regression Results}
\begin{tabular}{lcccc}
\hline
\textbf{Features} & \textbf{Coef} & \textbf{CI} & \textbf{P$>|z|$}  \\
\hline
\textbf{Intercept} & 0.16 & -0.27 – 0.59 & 0.469 \\
\textbf{Uptime} & -0.01 & -0.07 – 0.05 & 0.815 \\
\textbf{Discount register} & 0.01 & 0.01 – 0.02 & $<$0.001 \\
\textbf{Payment digital wallet} & 0.03 & -0.13 – 0.18 & 0.724 \\
\textbf{Price register} & -0.00 & -0.01 – 0.00 & 0.183 \\
\textbf{Free bulk search}  & -0.07 & -0.19 – 0.06 & 0.290 \\
\textbf{Payment crypto} & 0.23 & -0.07 – 0.52 & 0.130 \\
\textbf{API} & 0.12 & -0.08 – 0.32 & 0.227 \\
\textbf{Free DNS} & 0.15 & -0.27 – 0.57 & 0.473 \\
\textbf{Payment transfer} & -0.10 & -0.27 – 0.07 & 0.241 \\
\textbf{Free web host} & 0.20 & -0.20 – 0.60 & 0.325 \\
\textbf{Free SSL cert} & -0.24 & -0.63 – 0.15 & 0.221 \\
\textbf{EmailPhone validated} & 0.03 & -0.20 – 0.25 & 0.828 \\
\textbf{Restrictions} & -0.21 & -0.32 – -0.10 & $<$0.001 \\
\textbf{Prevention} & -0.13 & -0.40 – 0.14 & 0.356 \\
\hline
\end{tabular}
\vspace{0.05cm}

\textbf{Random Effects} \\
\begin{tabular}{lcc}
\hline
 & \textbf{Value} &  \\
\hline
$\sigma^2$ & 0.13 & \\
$\tau_{00~\text{TLD}}$ & 0.03 & \\
$\tau_{00~\text{Registrar}}$ & 0.07 & \\
ICC & 0.41 & \\
$N_{Registrar}$ & 38 & \\
$N_{TLD}$ & 293 & \\
Observations & 29890 & \\
Marginal $R^2$ / Conditional $R^2$ & 0.109 / 0.474 & \\
\hline
\end{tabular}
    \label{tab:rcoeffs}
\end{table}

\begin{figure}
  \centering 
  \subfloat{\includegraphics[width=0.41\textwidth]{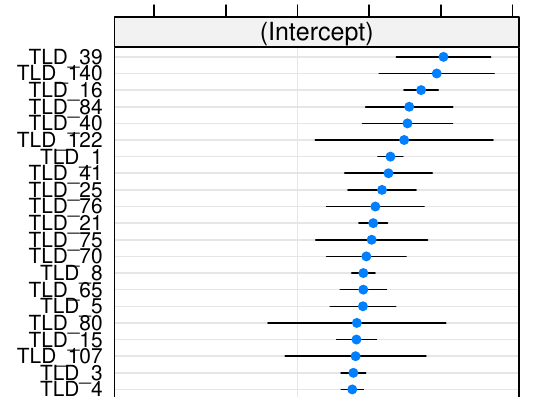}} 
  \qquad 
  \subfloat{\includegraphics[width=0.4\textwidth]{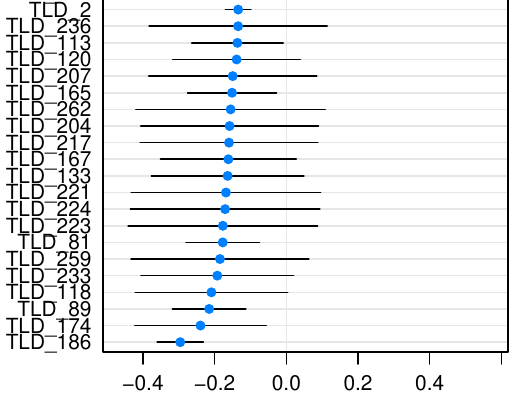}} 
  \caption{Random effects at the TLD level.}
  \label{fig:figureTLDs}
\end{figure}

\end{document}